\pdfoutput=1
\documentclass[british,a4paper,12pt]{phdthesis}
\usepackage[latin1]{inputenc}

\ifx\undefined\pdfoutput
\usepackage[dvips]{graphicx}
\usepackage[dvips]{color}
\DeclareGraphicsExtensions{.eps}
\else
\usepackage[pdftex]{graphicx}
\usepackage[pdftex]{color}
\DeclareGraphicsExtensions{.png,.pdf}
\fi
\usepackage[british]{babel}
\usepackage[english,vario]{fancyref}
\usepackage{setspace}
\usepackage{url}
\usepackage{amsmath}
\usepackage{enumerate}
\usepackage{appendix}

\makeatletter
 \renewcommand*{\@fancyref@page@ref}{%
  \let\vref@space\space
  \@ifnextchar[
  \@vpageref{\@vpageref[\unskip]}%
}%
\makeatother

\title{Enabling the Remote Acquisition of Digital Forensic Evidence through Secure Data Transmission and Verification}

\author{Mark Scanlon}

\authoraddress{
           School of Computer Science and Informatics,\\
           University College Dublin, Belfield, Dublin 4, Ireland.
     }

\date{\today}

\submission{A Thesis submitted to the\\
  National University of Ireland, Dublin\\
  for the degree of M. Sc.\\
  in the\\ College of Engineering, Mathematical and Physical Sciences}

\department{School of Computer Science and Informatics} \hod{J.
Carthy, Ph. D.} \supervision{M. Tahar Kechadi, Ph. D.}
\submissionmonth{September} \submissionyear{2009}

\begin{document}

\maketitle \tableofcontents \listoftables \listoffigures

\begin{abstract}
 \vspace{-1.5cm}
Providing the ability to any law enforcement officer to remotely transfer an image from any suspect computer directly to a forensic laboratory for analysis, can only help to greatly reduce the time wasted by forensic investigators in conducting on-site collection of computer equipment. RAFT (Remote Acquisition Forensic Tool) is a system designed to facilitate forensic investigators by remotely gathering digital evidence. This is achieved through the implementation of a secure, verifiable client/server imaging architecture. The RAFT system is designed to be relatively easy to use, requiring minimal technical knowledge on behalf of the user. One of the key focuses of RAFT is to ensure that the evidence it gathers remotely is court admissible. This is achieved by ensuring that the image taken using RAFT is verified to be identical to the original evidence on a suspect computer.
\end{abstract}

\begin{dedication}
  \vspace{-1.5cm}

This thesis is dedicated to my parents, Philomena and Larry Scanlon, who have always been there for me. This thesis is also dedicated to my girlfriend, Joanne Conway, who has supported, encouraged and motivated me throughout the last five years and especially throughout my research.

\end{dedication}

\begin{acknowledgements}
  \vspace{-1.5cm}
With no doubt, the work on this thesis has been the most challenging endeavour I have undertaken so far. I am thankful to my supervisor, Prof. M-Tahar Kechadi, for his guidance and encouragement. I would like to thank the staff in the School of Computer Science and Informatics, University College Dublin (UCD) for providing me with the opportunity to learn, facilities to perform my research, and a motivating environment that carried me forward through my course work. 

I would like to specifically thank all involved in the Centre for Cybercrime Investigation. My gratitude goes to my friends Cormac Phelan, Alan Hannaway, Damir Kahved\v zi\'c, John-Michael Harkness and Dr. Pavel Gladyshev for many interesting and developing discussions, presentations and collaborations. Many thanks to all my immediate friends for their constant encouragement and support.

\end{acknowledgements}

\begin{listofpublications}
\begin{itemize}


\item Conference Papers\\\\
M. Scanlon and M-T. Kechadi, ``Online Acquisition of Digital Forensic Evidence'', In the International Conference on Digital Forensics and Cyber Crime (ICDF2C 2009), Albany, New York, USA, September 30 - October 2, 2009.

\item Conference Presentations\\\\
M. Scanlon, ``RAFT - Remote Acquisition Forensic Tool'', 5$^{th}$ Irish Criminology Conference, UCD Institute of Criminology, Dublin, Ireland, June 15, 2009.

M. Scanlon, ``RAFT - Remote Acquisition Forensic Tool'', In the Cyber Terrorism and Crime Conference (CYTER 2009), Prague, Czech Republic, June 22-24, 2009.

\end{itemize}
\end{listofpublications}
\onehalfspacing


\chapter{Introduction}
\label{ch:introduction}

\section{Background}
\label{ch2:background}
Current trends in technology are putting computers with high-bandwidth Internet connections into the hands of regular criminals. As this phenomenon continues, an increasing number of traditional crimes are being aided by computers, e.g., fraud, identity theft, phishing, terrorism, online child sexual exploitation etc. As a result, digital forensic investigators are becoming overwhelmed with the number of cases they have to deal with. Traditional digital forensic investigations commence with the investigators leaving their laboratory to visit the crime scene, where they collect all the relevant evidence, and bring it back to the forensic laboratory for secure storage and analysis. This evidence may then lay untouched for extended periods while the investigating team deals with the backlog of cases.

If regular law enforcement officers had the ability to remotely transfer an image from any suspect computer directly to a forensic lab, it would help to significantly reduce the investigation time used to conduct on-site collection of computer equipment. The current approach for acquiring digital evidence is based on impounding the suspect computer system and examining it in a forensics laboratory. This examination is usually carried out on a copy of the original data \cite{gao}. In this thesis, we introduce a solution to reduce the time taken to acquire the necessary evidence. We propose RAFT (Remote Acquisition Forensic Tool), a remote forensic hard drive imaging tool, that is designed to boot off a Linux Live CD or USB memory stick. The evidence acquisition is achieved through the implementation of a secure, verifiable client/server imaging architecture. The suspect computer is booted using a customised Linux Live distribution and any hard drives or removable media connected to the computer are able to be securely imaged over an Internet connection directly to the RAFT Server. This system is designed to equip any law enforcement or investigating officers with the ability to easily perform digital evidence acquisition, which would traditionally require the expertise of an on-site forensic investigator. One key objective of RAFT is to ensure that any evidence gathered is court admissible. RAFT achieved this by ensuring that any interactions the system has with the evidence are conducted in a forensically sound, reliable and reproducible manner and that the imaged evidence taken is forensically verified to be identical to the original evidence.

\section{Primary Objectives}
\label{chap:Objectives-1} 
Many of existing digital evidence capturing tools are designed for use in a forensic laboratory. These traditional digital forensic tools are all reliant in the suspect computer equipment being seized and brought to the laboratory prior to imaging the hard drives and collecting the evidence. The motivation for the research detailed in this thesis is to equip regular law enforcement officers with the ability to remotely collect digital forensic evidence directly at the crime scene without requiring the presence of a digital forensic investigator. The primary objectives of this research are as follows:
\begin{enumerate}
\item Provide an insight into the technical requirements of the design and implementation of a remote forensic evidence acquisition tool and the transmission of large volumes of digital forensic evidence over the Internet.
\item Show the application of a remote digital evidence capturing system as a plausible option for forensic investigation.
\item Design an architecture for a remote digital forensic evidence capturing system. Such a system should be forensically verifiable, cost effective, expandable, reliable and widely compatible with current computer hardware. This system should improve on the time taken to get digital evidence into an ``investigation ready'' state in the forensic laboratory, while producing reliable, reproducible results.
\item Prototype the system and perform experimental analysis to measure the viability of the system.
\item Present the performance results achieved from testing the system.
\item Present applications and future potential applications of the system designed.
\end{enumerate}

\section{Contribution of This Work}
\label{sec:Contribution-of-this-work}
Many of the tools available in the field of digital evidence acquisition are based upon the imaging and analysis taking place in a forensic laboratory, e.g., EnCase (Section \ref{ch2:encase}), Forensic Toolkit (Section \ref{ch2:ftk}), FRED (Section \ref{ch2:fred}), etc. ``Typically, only a small fraction of the examined data is of interest in an investigation (e.g., one or two rogue machines out of tens, or hundreds). Thus, a lot of the effort in copying and carefully examining a large number of targets will be in vain" \cite{gao}. This existing research is concentrated around the procedures that should be implemented after the physical confiscation of the computer equipment. The research outlined as part of this thesis results in relevant evidence being in an ``investigation-ready" state as early into the investigation as possible. The contribution of this research can be summarised with the following points:

\begin{itemize}
\item Design of a forensically sound remote forensic digital evidence acquisition system which will result in the collection of court-admissible evidence. This system enables any law enforcement officer to capture digital evidence and for this evidence to be sent directly to a server in the forensic laboratory. This results in the forensic investigators being able to spend more time in the laboratory analysing evidence, as opposed to being in the field collecting it. This design extends to defining how to best deal with the issues of cost, speed, compatibility and redundancy of the system while ensuring that the process is reproducible and reliable.
\item Proof of the viability of the system through experimentation of all the necessary components. Each component in the system was individually tested to ensure the forensic integrity of the data collected.
\item Performance results from testing ``real-world" scenarios where such a system may be used, i.e., collecting evidence from a suspect computer in residential environment and from a corporate computer in an enterprise environment.
\end{itemize}

\section{Structure of This work}
\label{sec:Structure-of-this-work}
This thesis is organised as follows: 

\begin{itemize}
 \item After introducing the context and highlighting the main goals of the project in Chapter \ref{ch:introduction}, in Chapter \ref{ch:LiteratureReview} we present a literature review of related research work and software tools relevant to the area of acquiring digital forensic evidence remotely. This chapter outlines some of the tools, systems, architectures, storage formats, and best practices associated with the field of digital forensics from a technical, cryptographical and legal perspective.
 
 \item Chapter \ref{ch:arch} presents the architecture and design of the remote digital evidence capturing system developed as part of this research. We also outline the design goals which should be incorporated into tools of this nature. A proof of concept, the Remote Acquisition Forensic Tool (RAFT), is presented whereby an implementation of the system outlined was built to the specifications outlined.

 \item Chapter \ref{ch:results} presents the results of comprehensive experiments carried out to prove the viability of such a system. This involved testing each component of the system to ensure the forensic integrity of the evidence collected. Any minor change resulting from the handling of the source devices and their digital copies, could result in any evidence discovered on these devices being rendering inadmissible to the court. This chapter also outlines the results of experiments conducted using the prototyped forensic evidence acquisition tool in ``real-world" scenarios.

 \item Chapter \ref{ch:conclusion} summarises and concludes this research. This chapter also outlines scenarios where the technology developed can be adapted and reused for additional purposes. Guidelines for further developments to this tool are also outlined and discussed.
\end{itemize}

\chapter{Literature Review}
\label{ch:LiteratureReview}

\section{Introduction}
\label{ch2:int}

This chapter outlines some of the digital evidence acquisition and investigation software and hardware tools commonly used by forensic investigators in law enforcement and private investigations such as EnCase, Forensic ToolKit and the Forensic Recovery of Evidence Device (FRED). Current research and open-source tools are outlined specifying their benefits and designs, e.g., Bluepipe, DCFLDD. Common digital evidence storage formats are also discussed, outlining the cross-compatibility between the tools available and the associated formats. Best practices associated with the field of digital forensics from a technical, cryptographical and legal perspective are discussed.

\section{Digital Forensic Investigation}
\label{ch2:evidence}
Generally speaking, the goal of a digital forensic investigation is to identify digital evidence relative to a specific cybercrime. Investigations rarely rely entirely on digital evidence to prosecute the offender, instead relying on a case built from physical evidence, digital evidence, witness testimony and cross-examination. However, when dealing solely with digital evidence, there are three major phases \cite{carrier-open}:

\begin{enumerate}
    \item \emph{Acquisition Phase} -- The acquisition phase is concerned with capturing the state of a digital system for later analysis. This is similar to the collection of physical evidence from a crime scene, e.g., taking photographs, collecting fingerprints, fibres, blood samples, tire patterns, etc. During this phase, it is generally very difficult to tell which evidence is relevant to the case, so the goal of this phase is to save all possible digital values (including all allocated and unallocated space on any storage device). 
    
    \item \emph{Analysis Phase} -- After a successful and complete acquisition of the system state from a suspect computer, the data acquired needs to be analysed to identify pieces of evidence. The analysis of evidence is carried out on an exact copy of the original evidence. This copy is verified against the original through the use of a hashing algorithm, as outlined in more detail in section \ref{ch2:hashfunctions}. Carrier \cite{carrier-open} defines three major categories of evidence a digital investigator needs to discover when conducting his analysis:
    \begin{itemize}
        \item Inculpatory Evidence -- This is any evidence which supports a given theory.
        \item Exculpatory Evidence -- This is any evidence which contradicts a given theory.
        \item Evidence of Tampering -- This is any evidence which cannot be related to any theory currently under investigation, but shows that the system was tampered with to avoid identification.
    \end{itemize}
    The procedure followed during this phase includes examining file and directory contents (including recovered deleted content) to draw verifiable conclusions based on any evidence that was found.
    \item \emph{Presentation Phase} -- The steps performed in the previous two steps are the same regardless of the type of investigation being conducted, e.g., corporate, law enforcement or military. However, the presentation phase will be different depending on corporate policy or local law. This phase presents the conclusions and their corresponding evidence that the digital investigator has deduced. In a court settings, the lawyers must first evaluate the evidence to confirm that it is court admissible.
\end{enumerate}

\section{Digital Forensic Software Tools}
\label{ch2:existing}

While the area of Computer Forensics and Cybercrime Investigation is a relatively new evolution of the more traditional computer security model, there are a small number of companies, research projects and open-source tools dedicated to aid the forensic investigator in conducting the acquisition and analysis of forensic evidence. A number of these tools are discussed in the following subsections:

\subsection{Bluepipe}
\label{ch2:bluepipe}

Bluepipe is a live digital forensic system for *NIX platforms that is a viable alternative to the traditional post-mortem analysis created by Y. Gao, G.G. Richard III and V. Roussev \cite{battistoni}. The creators published a paper in the International Journal of Digital Evidence entitled ``Bluepipe: A Scalable Architecture for On-the-Spot Digital Forensics" \cite{gao}. In this paper, the authors presented the Bluepipe architecture and the Bluepipe remote forensic protocol. The Bluepipe architecture was designed to counteract the following issues with the traditional approach to forensic evidence acquisition:

\begin{enumerate}
\item How invasive the impounding of suspect computers can be to an operational business. If the collection of evidence is as non-invasive as possible, i.e., not requiring the business to shut down and part with operational equipment, it is more likely that a business will willingly cooperate with the investigation, assuming no warrant has been granted.
\item Some businesses could have sensitive information stored on their computer equipment which is of a very high level of confidentiality and potentially irrelevant to the investigation being processed outside of the company.
\item The logistical implications of moving and storing a large volume of computer equipment from a company's premises to the forensics laboratory.
\item Not all the collected information is relevant to the investigation and usually the relevant evidence is a very small proportion of the total collected.
\end{enumerate}

Bluepipe is designed on a client/server paradigm. The server runs on the target machine, while the forensic investigator runs the client. The server's responsibility is to execute any commands received from the client through the Bluepipe protocol. The communication between the client and server is supported through Bluetooth and 802.11 wireless networking. The Bluepipe protocol is capable of sending simple XML based commands to the target machines such as locate, get, grep, listdir, listpartitions and hash. These commands enable the investigator to browse around the target machine remotely and attempt to discover any evidence relevant to the case he is investigating.



\subsection{DCFLDD}
\label{ch2:dcfldd}

DCFLDD is a digital forensic imaging tool created by the Department of Defence Computer Laboratory in the United States \cite{dcfldd}. It is an enhanced version of the GNU ``dd'' *NIX command included in a collection of tools called Forensic and Incident Response Environment (FIRE). DCFLDD is a natural progression of the ``dd'' command with additional forensic oriented features such as \cite{forte2006state}:
\begin{itemize}
\item \emph{On-the-fly Hashing} -- While the input is being copied, it is possible for a configurable hash value to be simultaneously computed. The dcfldd command is capable of calculating most common hash functions, as outlined in \ref{ch2:hashoverview}.
\item \emph{Split Output} -- The tool is capable of splitting the output at a defined size. This is especially a useful option if the resultant collected data needs to be stored on a fixed size medium, e.g., a CD or DVD.
\item \emph{Piped Output} -- The data collected through the dcfldd command is capable of being output to any other program by way of a pipe.
\item \emph{Logging} -- The tool is capable of creating log files with hash values, input and output locations, disk sizes and other useful information.
\end{itemize}

The evidence captured using DCFLDD is stored in the ``raw format", outlined in \ref{ch2:raw}, and as a result it is compatible with any of the analysis tools on the market (both open source and commercial). It is possible to configure the dcfldd command with command line arguments to save the hashes of all collected files into a separate file. A sample usage of the dcfldd command is shown below \cite{forte2006state}:
\begin{verbatim}
dcfldd -hashwindow=BYTE -hash=sha512 -hashlog=FILE 
if=DISK1 of=DISK2
\end{verbatim}
An explanation of the command lines arguments outlined is included below:
\begin{itemize}
\item \emph{hashwindow} -- This argument specifies how often the program should hash the data copied, i.e., after how many bytes should each hash be produced. It's possible to send a single copy of a file and create many hash values for each sequential chunk of the file.
\item \emph{hash} -- This specifies which hash function to use. Acceptable values are md5, sha1, sha256, sha384 or sha512. Multiple hash sums can be computed simultaneously by including the required hash names in a comma separated list, e.g., md5,sha384.
\item \emph{hashlog} -- Send the hash sums produced to a file as opposed to the stderr.
\item \emph{if (Input File)} -- Read from a file instead of stdin. The direct path to a partition or a hard disk can be passed here as an argument, e.g., \begin{verbatim}if=/dev/sda3\end{verbatim}
\item \emph{of (Output File)} -- Write to a file instead of stdout. Multiple output files can be specified for the data being copied to be written to multiple locations simultaneously.
\end{itemize}

\subsection{EnCase}
\label{ch2:encase}

\begin{figure}
\centering
\includegraphics[width=0.98\textwidth]{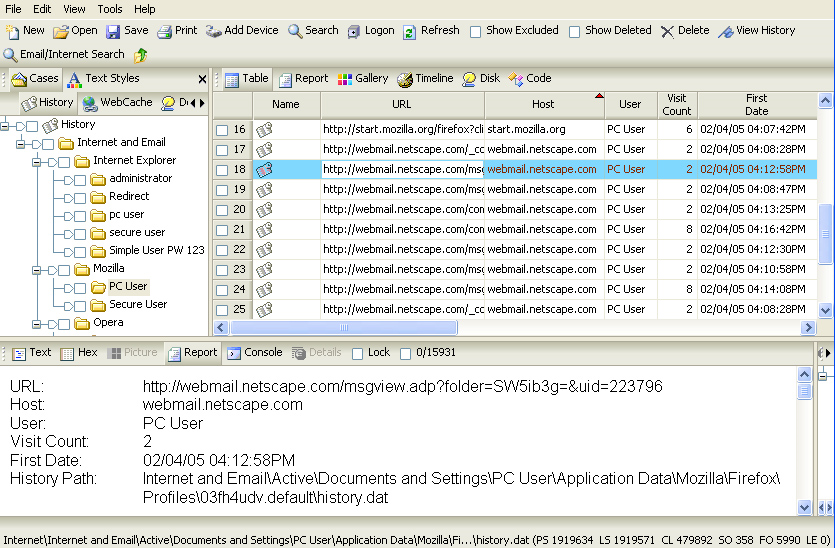}
\caption[EnCase Forensic screenshot showing internet history]{Screenshot of EnCase Forensic showing its analysis of the internet history of a collected hard drive image. EnCase collates the history from Internet Explorer, Mozilla, Opera and Macintosh Safari \cite{guidance}.}
\label{fig:encasehistory}
\end{figure}

EnCase is a forensics image acquisition, analysis and reporting tool created by Guidance Software \cite{guidance}. By many professionals, it is seen as the de facto standard for digital investigations. Over 2000 law enforcement agencies worldwide use EnCase according to Jennifer Higdon, spokesperson for Guidance Software \cite{garber2001encase}. The EnCase Forensic tool is capable of performing a multitude of tasks for the investigator, such as \cite{guidance}:

\begin{enumerate}
\item \emph{Evidence Acquisition} -- Acquisition of data from a locally connected hard disk. The hard disk is generally connected to the forensic workstation through a write-blocker.
\item \emph{Automated Tools} -- These automated processes operate after the user configures a range of filters to examine a drive. For example, the highlighting of all digital images/photographs, recovery of any deleted partitions and extraction of the registry information is all automated.
\item \emph{Analysis} -- EnCase is capable of quickly displaying, searching and parsing over 400 file formats, any internet, email and instant messaging histories (as is shown in the screenshot in Fig. \ref{fig:encasehistory}), and numerous file system formats. The investigator has the ability to bookmark any discovered evidence for later ease of access and reporting.
\item \emph{Reporting} -- A number of automatic reports are available including a list of all files and folders, detailed web history, log records, hard drive information (including partition information) and bookmarked files and images.
\end{enumerate}

\subsection{Forensic Toolkit}
\label{ch2:ftk}

\begin{figure}
\centering
\includegraphics[width=0.98\textwidth]{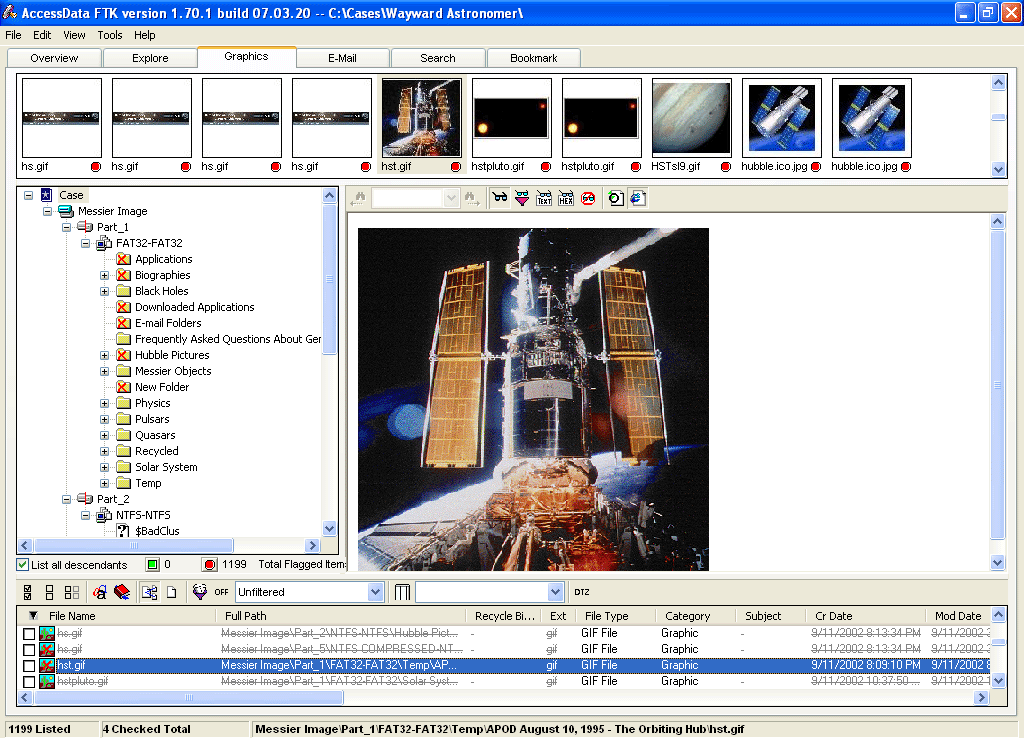}
\caption[Forensic ToolKit screenshot showing automated category allocation]{Screenshot of Access Data's Forensic ToolKit (FTK) showing the categories of automatically collected files.}
\label{fig:ftk}
\end{figure}

Forensic Toolkit (FTK) is a tool created by Access Data \cite{ftk}. FTK has similar functionality to EnCase, i.e., it is an ``all-in-one" image acquisition, analysis and reporting tool with the ability to automate common investigative tasks. As can be seen in Fig. \ref{fig:ftk}, FTK is capable of the automatic organisation of data into categories (as can be seen with the tabs along the top of the window). A notable feature of FTK is that it has the capability to use a database-driven architecture to keep track of the analysis of a given disk for distributed analysis. This distributed analysis is used for automated data pre-processing, e.g., recovering deleted files and partitions, structuring files into categories etc. It also includes ``Password Recovery Toolkit" and ``Distributed Network Attack". Password Recovery Toolkit enables the investigators to crack the password of over 80 different applications using brute force methods. The Distributed Network Attack utilises multiple computers in a distributed system to perform dictionary attacks on encrypted or password protected files.

\begin{figure}
\centering
\includegraphics[width=0.98\textwidth]{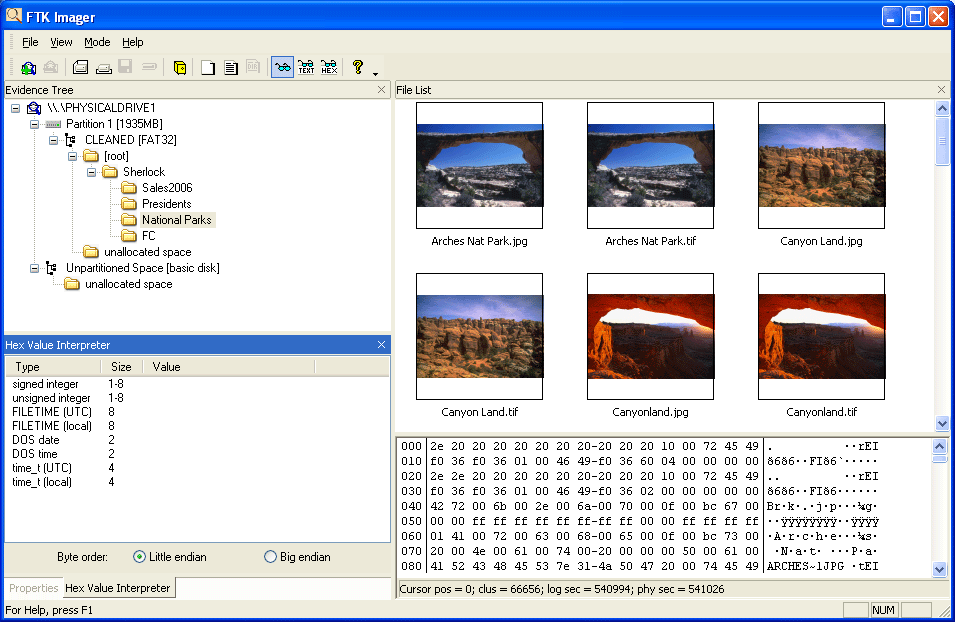}
\caption[FTK Imager screenshot displaying image directory preview]{Screenshot of FTK Imager showing its image directory preview. FTK Imager allows the investigator to examine the live contents of a hard drive (both in the allocated and unallocated space) and allows the investigator to take a forensically sound image of the disk \cite{ftk}.}
\label{fig:ftkimager}
\end{figure}

FTK Imager is the evidence capturing tool packaged along with the Forensic Toolkit. As can be seen in Fig. \ref{fig:ftkimager}, the tool is capable of showing a live preview of the content on a suspect disk. This could positively influence the investigation by targeting the analysis at the relevant areas. The FTK Imager is capable of producing images in the raw data format, as outlined in section \ref{ch2:raw}, \cite{dfrws2006}.

\section{Digital Forensic Hardware}
\label{ch2:enterprisehardware}

The current standard hardware device used for digital evidence acquisition in the forensic laboratory is the FRED workstation, as outlined in section \ref{ch2:fred} below, and the portable FRED called the FREDDIE, as outlined in section{\ref{ch2:freddie}. Both of these hardware devices are bundled with a number of write-blocked connections for acquiring evidence from common storage devices.

\subsection{Forensic Recovery of Evidence Device}
\label{ch2:fred}

The Forensic Recovery of Evidence Device (FRED) is a collection of equipment tailored for digital investigations available from Digital Intelligence \cite{fred}. Fig. \ref{fig:fred} shows the FRED workstation with the write blocked disk connection ports on the front of the device. The shelf at the front of the device is non-conductive and contains fans to cool the disk during evidence acquisition. Each FRED contains a collection of write-blocked (read-only) ports including SATA, IDE, SCSI, USB and FireWire. The FRED system is designed to aid digital investigators by ensuring that all interactions with the original source of evidence is read-only. FREDs are still reliant on a standard operating system, e.g., a Windows variant, and one or more of the software acquisition and analysis tools described above.

\begin{figure}
\centering
\includegraphics[width=0.5\textwidth]{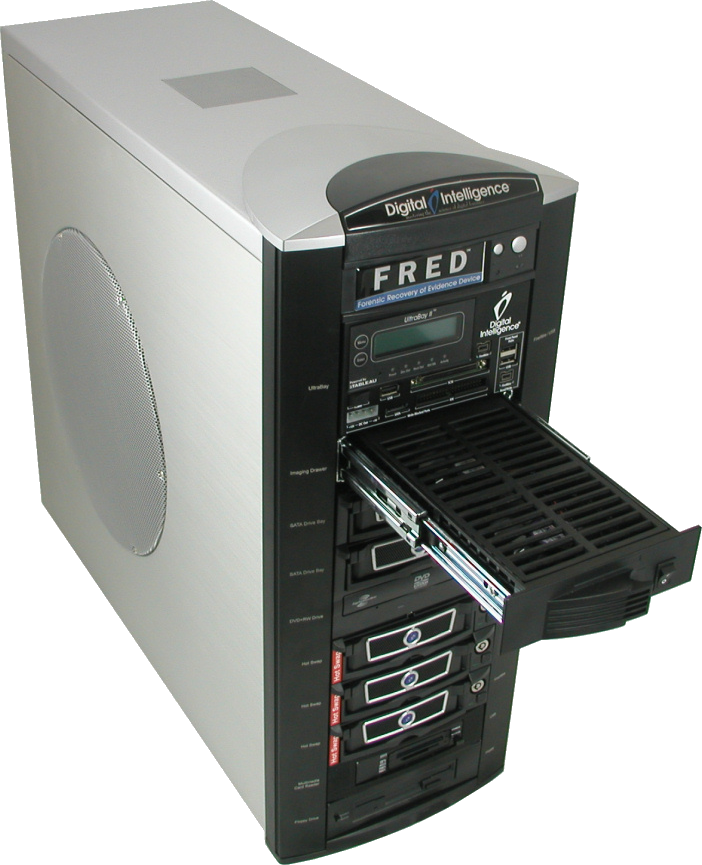}
\caption[Forensic Recovery of Evidence Device]{Forensic Recovery of Evidence Device from Digital Intelligence \cite{fred}}
\label{fig:fred}
\end{figure}

It is the only hardware tool with a write blocked FireWire port. This enables it to acquire evidence from Apple Mac computers booted into FireWire mode, i.e., giving direct access to the hard disk through the FireWire port without booting the full operating system. FRED systems are also capable of acquiring write-blocked data from floppies, CDs, DVDs and a variety of memory card formats such as Compact Flash, Micro Drives, Smart Media, Memory Stick, Memory Stick Pro, xD Cards, Secure Digital Media and Multimedia Cards.

\subsection{Forensic Recovery of Evidence Device Diminutive Interrogation Equipment}
\label{ch2:freddie}

Forensic Recovery of Evidence Device Diminutive Interrogation Equipment (FREDDIE) is the portable configuration of FRED \cite{freddie}. It is capable of much of the same functionality of FRED device, with the only downsides being that due to its portable design, there is a physical hard disk limitation as to how many different sources of evidence you are able to capture during the same excursion. This limitation is easily avoided when using the full sized FRED, as it is generally connected to a large network attached storage device.

\begin{figure}
\centering
\includegraphics[width=0.5\textwidth]{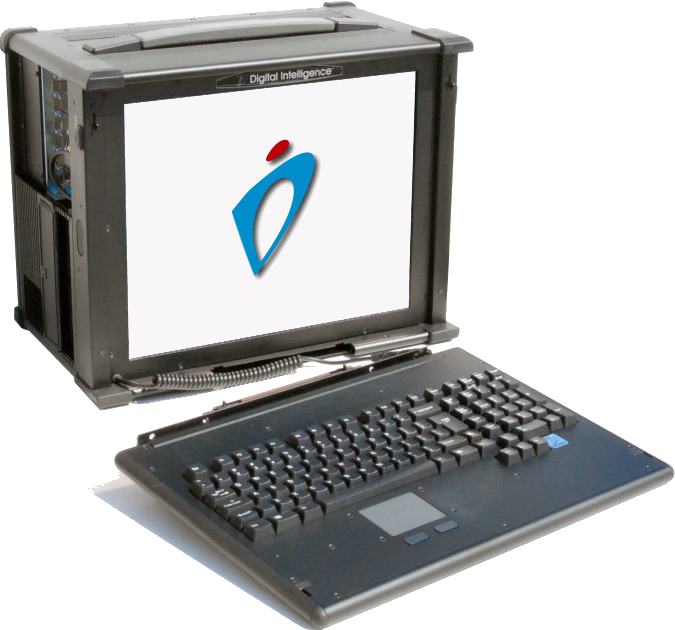}
\caption[Forensic Recovery of Evidence Device Diminutive Interrogation]{Forensic Recovery of Evidence Device Diminutive Interrogation Equipment from Digital Intelligence \cite{freddie}}
\label{fig:freddie}
\end{figure}

\section{Evidence Storage Formats}
\label{ch2:storage}

There is currently no universal standard for the format that digital evidence and any case related information is stored. This is due to the fact that there are no state or international governmental policies to outline a universal format. Many of the vendors developing forensic tools have developed their own proprietary format. With such a relatively small target market, it sometimes makes business sense for them to try and lock their customers into buying only their software in the future. There have been a small number of attempts at creating open formats to store evidence and any related metadata. This section describes the most common of these formats below.

\subsection{Common Digital Evidence Storage Format}
\label{ch2:cdesf}

The Common Digital Evidence Storage Format (CDESF) Working Group was created as part of the Digital Forensic Research Workshop (DRFWS) in 2006. The goal of this group was to create an open data format for storing digital forensic evidence and associated metadata from multiple sources, e.g., computer hard drives, mobile Internet devices, etc. \cite{cdesf}. The format which the CDESF working group were attempting to create would have specified metadata capable of storing case-specific information such as case number, digital photographs of any physical evidence collected and the name of the digital investigator conducting the investigation. In 2006, the working group produced a paper outlining the advantages and disadvantages of various evidence storage formats \cite{dfrws2006}.

Ultimately due to resource restrictions, the CDESF working group was disbanded in 2007 before accomplishing their initial goal.

\subsection{Raw Format}
\label{ch2:raw}

According to the CDESF Working Group, ``the current de facto standard for storing information copied from a disk drive or memory stick is the so-called ``raw" format: a sector-by-sector copy of the data on the device to a file" \cite{evidencestandards}. The raw format is so-called due to the fact that it is simply a file containing the exact sector-by-sector copy of the original evidence, e.g., files, hard disk/flash memory sectors, network packets, etc. Raw files are not compressed in any manner and as a result, any deleted or partially overwritten evidence that may lay in the slackspace of a hard disk is maintained. All of the commercial digital evidence capturing tools available today have the capability of creating raw files. Digital evidence stored in the raw format is also compatible with all of the commercial digital investigation analysis tools outlined in \ref{ch2:existing}.

\subsection{Advanced Forensic Format}
\label{ch2:aaf}

The Advanced Forensic Format (AAF) is an open source, extensible format created by S. Garfinkel in Basis Technology in 2006 \cite{aaf}. The AAF format has a major emphasis on efficiency and as a result is partitioned into two layers; the disk representation layer which defines segment name used for storing all data associated with an image and the data storage layer which defines how the image is stored, be it binary or XML\cite{containers}. The format specifies three variants; AFF, AFD and AFM. AFF stores all data and metadata in a single file, AFD stores the data and metadata in multiple small files, and AFM stores the data in a raw format and the metadata is stored in a separate file \cite{dfrws2006}. \cite{containers}

\subsection{Generic Forensic Zip}
\label{ch2:gfzip}

Generic Forensic Zip (gfzip) is an open source project to create a forensically sound compressed digital evidence format based on AAF \ref{ch2:aaf} \cite{gfzip}. Due to the fact that it is based upon the AAF format, there is limited compatibility between the two in terms of segment based layout. One key advantage that gfzip has over the AAF format is that gfzip seeks to maintain compatibility with the raw format \ref{ch2:raw}. It achieves this by allowing the raw data to be placed first in the compressed image \cite{containers}.

\subsection{Digital Evidence Bag (QinetiQ)}
\label{ch2:debq}

The method for traditional evidence acquisition involves a law enforcement officer collecting any relevant items at the crimescene and storing the evidence in bags and seals. These evidence bags may then be tagged with any relevant case specific information, such as \cite{debq}:
\begin{itemize}
\item Investigating Agency / Police Force
\item Exhibit reference number
\item Property reference number
\item Case/Suspect name
\item Brief description of the item
\item Date and time the item was seized/produced
\item Location of where the item was seized/produced
\item Name of the person that is producing the item as evidence
\item Signature of the person that is producing the item
\item Incident/Crime reference number
\item Laboratory reference number
\end{itemize}

Physical evidence containers, such as evidence bags, are trusted due to the well understood and practised process called ``chain of custody" \cite{debwet}.

Digital Evidence Bag (DEB) is a digital version of the traditional evidence bag, created by Philip Turner in 2005 \cite{debq}. DEB is based on an adaptation of existing storage formats, with potentially infinite capacity. The data stored in a DEB is stored in multiple files, along with metadata containing the information that would traditionally be written on the outside of an evidence bag. There are currently no tools released that are compatible with the DEB format.

\subsection{Digital Evidence Bag (WetStone Technologies)}
\label{ch2:debwet}

In 2006, C. Hosmer, from WetStone Technologies Inc., published a paper outlining the design of a Digital Evidence Bag (DEB) format for storing digital evidence \cite{debwet}. This format for storing is independent from the Digital Evidence Bag outlined in \ref{ch2:debq}. The format emerged from a research project funded by the U.S. Air Force Research Laboratory. The motivation for this format was similar to the motivation for that described in \ref{ch2:debq}, i.e., to metaphorically mimic the plastic evidence bag used by crime scene investigators to collect physical evidence such as blood, fibres, hairs etc. This format will be released publicly when complete.

\subsection{EnCase Format}
\label{ch2:encaseformat}

The EnCase format for storing digital forensic is proprietary to the evidence analysis tool of the same name as outlined in section \ref{ch2:encase}. It is by far the most common evidence storage option used by law enforcement and private digital investigation companies \cite{containers}. Because of the proprietary nature of the format, along with the lack of any formal specification from Guidance Software \cite{guidance}, much remains unknown about the format itself. Some competitors to Guidance Software have attempted to reverse engineer the format to provide an element of cross-compatibility with their tools \cite{aaf}. EnCase stores a disk image as a series of unique compressed pages. Each page can be individually retrieved and decompressed in the investigative computer's memory as needed, allowing a somewhat random access to the contents of the image file. The EnCase format also has the ability to store metadata such as a case number and an investigator \cite{aaf}.

\section{Evidence Handling}
\label{ch2:handling}

When analysing physical evidence, the commonly used procedure is known as the ``chain of custody" \cite{debq}. The chain of custody commences at the crime scene where the evidence is collected, when the investigating officer collects any evidence he finds and places it into an evidence bag. This evidence bag will be sealed to avoid any contamination from external sources and signed by the officer and will detail some facts about the evidence, e.g., description of evidence, location it was found, date and time found etc. The chain of custody will then be updated again when the evidence is checked into the evidence store in the forensic laboratory. When it comes to analysing the evidence, it will be checked out to the analysts' custody and any modification to the evidence required to facilitate the investigation, e.g., taking a sample from a collected fibre to determine its origin or unique properties. Each interaction with the evidence will be logged and documented. 

The procedures outlined above for physical evidence need to be slightly modified for digital evidence acquisition and analysis. Due to the fact that digital evidence is analysed on forensic workstations, most of the above sequences can be automated into concise logging of all interactions. During a digital investigation, there is no requirement to modify the existing evidence in any way. This is because all analysis is conducted on an image of the original source and any discovered evidence can be extracted from this image, documented and stored separately to both the original source and the copied image. It is imperative when dealing with all types of evidence that all procedures used are reliable, reproducible and verifiable. In order for evidence to be court admissible, it must pass the legal criteria for the locality that the court case is being heard, as outlined in greater detail in section \ref{ch2:legal} below.

\subsection{What does ``Forensically Sound" really mean?}
\label{ch2:sound}

Many of the specifications for digital forensic acquisition tools, analysis tools, storage formats and hash functions state that the product in question is ``forensically sound" or that the product works with the digital evidence in a ``forensically sound manner", without specifying exactly what the term means. In 2007, E. Casey published a paper in the Digital Investigation Journal entitled ``What does ``forensically sound" really mean ?" \cite{forensicallysound}. 

In this paper, Casey outlines some of the common views of forensic professionals regarding dealing with digital forensic evidence. Purists state that any digital forensic tools should not alter the original evidence in any way. Others point out that the act of preserving certain types of evidence necessarily alters the original, e.g., a live memory evidence acquisition tool must be loaded into memory (altering the state of the volatile memory and possibly overwriting some latent evidence) in order to run the tool and capture any evidence contained in the memory. Casey then goes onto to explain how some traditional forensic process require the altering of some of the evidence in order to collect the required information. For example, collecting DNA evidence requires taking a sample from some collected evidence, e.g., a hair. Subsequently, the forensic analysis of this evidentiary sample (DNA profiling) is destructive in its nature which further alters the original evidence.

Casey summarises that from a forensic standpoint, evidence acquisition and handling should modify the evidence as little as possible and when modification is unavoidable, it should be well documented and considered in the final analytical results. ``Provided the acquisition process preserves a complete and accurate representation of the original data, and its authenticity and integrity can be validated, it is generally considered forensically sound" \cite{forensicallysound}.

\subsection{Splitting Evidence}
\label{ch2:splitting}

It is not always possible to store the entire image of a particular storage device in one large file. This could be for a number of reasons, such as the evidence being stored on a FAT32 formatted hard drive which is only capable of addressing a file less than $2^{32}$ bytes (4,294,967,296 bytes or 4 gigabytes) or if evidence needs to be backed up to external media, e.g., a data CD or DVD, capable of storing 700MB and 4.7GB respectively. If this evidence is going to be transmitted over the Internet, it should be a requirement of any such system to split the evidence into smaller parts to minimise the cost of dropped connections. The CDESF working group conducted a survey in 2006 and found that each of the evidence storage formats they tested was capable of allowing split archiving and storage of evidence \cite{dfrws2006}. 

Should any tool split the evidence during acquisition, for transmission or storage purposes, this collected evidence should be recompilable into the original source for examination purposes. To ensure forensic integrity, the tools used for splitting and recompiling the evidence should be able to verify the recompiled image against the original untouched source.

\subsection{Compressing Forensic Evidence}
\label{ch2:compression}

Compressing digital forensic evidence into common compressed formats, e.g., *.zip, *.rar, *.7z, *.tar.gz, etc., would result in an unrecoverable loss of evidence. When evidentiary data is compressed it is imperative for only uninitiated space on a disk to be excluded from the compressed file. If compressing an entire hard drive, each of the aforementioned compression functions will not include any slackspace on the drive. This slackspace can contain files and evidence that the suspect user has deleted from his computer, assuming that the particular sectors that were used to store the file have not already been overwritten. There are some evidence storage formats, as outlined in section \ref{ch2:storage}, which compress the data collected in a forensically sound manner.

\section{Cryptographic Hash Functions}
\label{ch2:hashfunctions}

Cryptographic hash functions are deterministic procedures which operate by taking in a block of data and produce a fixed length digital fingerprint or cryptographic hash value/sum. The data input to a hash function is commonly referred to as the ``message``, while the hash sum produced is referred to as the digest.

The ideal collision resistant cryptographic hash function (h) has four main properties, defined by B. Preneel as part of his Ph.D. thesis in 1993 \cite{chf}:
\begin{enumerate}
\item The description of h must be publicly known and should not require any secret information for its operation.
\item The argument/message X can be of arbitrary length and the result h(X) has a fixed length of n bits (with n $\geq$ 128).
\item Given h and X, the computation of h(X) must be ``easy".
\item The hash function must be one-way in the sense that given a Y in the image of h, it is infeasible to find a message X such that h(X) = Y , i.e., it should be impractical to modify a message without changing its hash. It should also be infeasible given X and h(X) to find a message X' $\neq$ X such that h(X') = h(X), i.e., finding two different messages with the same hash should be unattainable.
\item The hash function must be collision resistant: this means that one should not find two distinct messages that hash to the same result. It also should not be feasible to find a message X that has a given hash sum h(X).
\end{enumerate}


\subsection{Collision Resistance}
\label{ch2:collisionresistance}

The measure of the unlikelihood of two different inputs to a hashing function returning the same hash sum is known as the collision resistance of the hash function. Generally speaking, the larger the internal state size that the hashing function has to operate with, the better the collision resistance of that function.


In 2005, Wang and Yu published a paper outlining their attempts to break a number of specified hash functions, entitled ``How to Break MD5 and Other Hash Functions" \cite{wangmd5}. In this paper they described a method for engineering two files which, when hashed using MD5, would result in having the same hash sum. In their experiments, they created two different files, F1 and F2, by reverse engineering them to have the specific bits in the specific file locations required for the hashing function to produce an identical hash sum. It is important to note that there currently is no documented evidence that, if given a specific file F1, that anyone is capable of engineering a second file F2 that has the same hash sum. As a result of this paper, the United States Computer Emergency Readiness Team (US-CERT), part of the United States' Department of Homeland Security,  published a vulnerability note stating that MD5 should be considered cryptographically broken and unsuitable for further use and that most United States governmental applications will be required to move to the SHA-2 family of hashing functions by 2010 \cite{certmd5}.

To date, no collisions have been found in any of the SHA-2 family of hashing functions.

\subsection{Avalanche Effect}
\label{ch2:avalanche}

The avalanche effect of a cryptographic hashing function refers to a desirable property whereby should the input file be modified slightly \cite{zhang}, e.g., changing a single bit of the file, the resultant hash sum produced changes significantly. The term ``avalanche effect" used to describe this property was created by H. Feistel in 1975 \cite{feistel}. Table 2.1 shows a sample set of common hashing functions along with sample hash sums they produce for two slightly different input files showing the influence the avalanche effect has on each function.

\begin{table}
\label{ch2:hashchoice}
\begin{center}
\begin{tabular}{| p{2.1cm} | p{1.2cm} | p{4.1cm} | p{4.1cm} | p{1cm} |}
\hline
Hash Algorithm  & Length in bits    & The quick brown fox jumps over the lazy dog & The quick brown fox jumps over the lazy cog & Diff \%\\
\hline
Adler32     & 32    & 
    \texttt{5BDC0FDA} & 
    \texttt{5BD90FD9} & 25.0\%\\
\hline
CRC32       & 32    & 
    \texttt{414FA339} & 
    \texttt{4400B5BC} & 87.5\%\\
\hline
Haval       & 128   & 
    \texttt{713502673D67E5FA 557629A71D331945} & 
    \texttt{4C9409BE8321D982 72D9252F610FBB5B} & 93.8\%\\
\hline
MD2     & 128   & 
    \texttt{03D85A0D629D2C44 2E987525319FC471} & 
    \texttt{6B890C9292668CDB BFDA00A4EBF31F05} & 93.8\%\\
\hline
MD4     & 128   & 
    \texttt{1BEE69A46BA81118 5C194762ABAEAE90} & 
    \texttt{B86E130CE7028DA5 9E672D56AD0113DF} & 93.8\%\\
\hline
MD5     & 128   & 
    \texttt{9E107D9D372BB682 6BD81D3542A419D6} & 
    \texttt{1055D3E698D289F2 AF8663725127BD4B} & 100\%\\
\hline
RipeMD128   & 128   & 
    \texttt{3FA9B57F053C053F BE2735B2380DB596} & 
    \texttt{3807AAAEC58FE336 733FA55ED13259D9} & 93.8\%\\
\hline
RipeMD160   & 160   & 
    \texttt{37F332F68DB77BD9 D7EDD4969571AD67 1CF9DD3B} & 
    \texttt{132072DF69093383 5EB8B6AD0B77E7B6 F14ACAD7} & 95.0\%\\
\hline
SHA-1       & 160   & 
    \texttt{2FD4E1C67A2D28FC ED849EE1BB76E739 1B93EB12} & 
    \texttt{DE9F2C7FD25E1B3A FAD3E85A0BD17D9B 100DB4B3} & 95.0\%\\
\hline
SHA-256     & 256   & 
    \texttt{D7A8FBB307D78094 69CA9ABCB0082E4F 8D5651E46D3CDB76 2D02D0BF37C9E592} & 
    \texttt{E4C4D8F3BF76B692 DE791A173E053211 50F7A345B46484FE 427F6ACC7ECC81BE} & 95.3\%\\
\hline
SHA-384     & 384   & 
    \texttt{CA737F1014A48F4C 0B6DD43CB177B0AF D9E5169367544C49 4011E3317DBF9A50 9CB1E5DC1E85A941 BBEE3D7F2AFBC9B1} & 
    \texttt{098CEA620B0978CA A5F0BEFBA6DDCF22 764BEA977E1C70B3 483EDFDF1DE25F4B 40D6CEA3CADF00F8 09D422FEB1F0161B} & 95.8\%\\
\hline
SHA-512     & 512   & 
    \texttt{07E547D9586F6A73 F73FBAC0435ED769 51218FB7D0C8D788 A309D785436BBB64 2E93A252A954F239 12547D1E8A3B5ED6 E1BFD7097821233F A0538F3DB854FEE6}& 
    \texttt{3EEEE1D0E11733EF 152A6C29503B3AE2 0C4F1F3CDA4CB26F 1BC1A41F91C7FE4A B3BD86494049E201 C4BD5155F31ECB7A 3C8606843C4CC8DF CAB7DA11C8AE5045} & 96.1\%\\
\hline

\end{tabular}
\end{center}
\caption[Example hash sums from popular hash functions]{Example hash sums for a small file containing the sentence outlined. The percentage difference shows the difference in the hash sums produced. While each character of a hash is hexadecimal, i.e., 1 of 16 possible values, it is notable that some hashing functions have differences greater than the expected maximum difference, i.e., $>$93.8\%. This is due to a more pronounced avalanche effect in the hashing function.}
\end{table}

\subsection{Overview of Common Hashing Algorithms}
\label{ch2:hashoverview}

While there are hundreds, if not thousands, of hashing functions in existence, the list of commonly used functions is significantly shorter. This is due to the fact that the National Institute for Standards and Technology (NIST) and the National Security Agency (NSA) in the United States have prioritised the standardisation of hashing functions. The most popular hashing functions, outlined below, are all based on the message digest principle. The message digest principle was designed by Ronald Rivest and constitutes a hash function taking in a message of arbitrary length and producing a fixed length message digest (hash value/sum) based on that input.

\subsubsection{MD Family}
\label{ch2:md}

The Message Digest algorithm family of hash functions were all created by Ronald Rivest, a professor in Massachusetts Institute of Technology, along with some collaboration from others. The family contains six iterations of the algorithms; MD, MD2 (1988), MD3 (1989), MD4(1990), MD5 (1991) and MD6 (2008.) From the original iteration up as far as MD5, the algorithms all produced 128-bit message digests. These MD hash values are expressed as 32 hexadecimal digits, as can be seen in table 2.1. MD6 is based on a variable length message digest size to improve performance for smaller inputs, and as a result the message digest can be anywhere in the range from 0 - 512 bits in length.

MD5 is a popular hash function used in numerous applications. Most of the tools available to the digital investigator rely on a combination of the CRC32 and the MD5 hash functions for maintaining data integrity \cite{dfrws2006}.

MD6 was entered into the competition for the SHA-3 Family of hash functions. However, in July 2009, the algorithm was withdrawn from the competition because in order for it to be fast enough to compete, the design would have had to compromise its resistance to differential attacks.

\subsubsection{SHA-0 and SHA-1 Family}
\label{ch2:sha1}

The first specification of the Secure Hashing Algorithm (SHA) family of hashing functions was published in 1993 by the US National Institute for Standards and Technology. This early specification is now known as the SHA-0 function. SHA-0 was withdrawn from use by the US National Security Agency in 1995 and was replaced by a modified version of the function; SHA-1. Both SHA-0 and SHA-1 produce 160-bit hash sums and they have a maximum input message size of $2^{64} - 1$ bits (or 2048 petabytes).

X. Wang, Y.L. Yin and H. Yu produced a paper entitled ``Finding Collisions in the Full SHA-1" in 2005 \cite{wangsha1}. This paper outlined the first attack on the SHA-1 hash function. The authors successfully found collisions on the SHA-1 function. They achieved this by first finding near-collisions. They then were able to discover full collisions based on the analysis of the near collisions. They conclude that although the SHA-1 family of hash functions has message expansion, it does not offer enough avalanche effect in terms of differing inputs.

\subsubsection{SHA-2 Family}
\label{ch2:sha2}

The SHA-2 Family consists of the following hash functions: SHA-224, SHA-256, SHA-384, and SHA-512. The number in the name of the hash function represents the output message digest size in bits. H. Gilbert and H. Handschuh produced a journal paper entitled ``Security Analysis of SHA-256 and Sisters" in 2004 \cite{gilbert} which published their results from the analysis of the SHA-2 family of hash functions. They found that the attacks that have broken the SHA-1 family no longer are applicable to the SHA-2 family.

The SHA-224 and SHA-256 have the same maximum input file size of $2^{64} - 1$ bits (or 2048 petabytes) as with the SHA-1 Family, while the SHA-384 and SHA-512 have a maximum of $2^{128} - 1$ bits (or 3.78 x $10^{22}$ petabytes).

\subsubsection{SHA-3 Family (in development)}
\label{ch2:sha3}

The United States National Institute of Standards and Technology (NIST), part of the Department of Commerce, are holding a development competition to decide on which hashing function to choose for the third iteration of the SHA Family. As part of the competition, NIST accepted over 50 entries into the first round of testing. This number was reduced down to 14 accepted into the second round which was announced in August 2009 \cite{sha3}. The remaining candidates in the second round are BLAKE, Blue Midnight Wish, CubeHash (Bernstein), ECHO (France Telecom), Fugue (IBM), Gr\o stl (Knudsen et al.), Hamsi, JH, Keccak (Keccak team, Daemen et al.), Luffa, Shabal, SHAvite-3, SIMD and Skein (Schneier et al.). The winner of the hashing function development competition and publication of the new SHA-3 standard are scheduled to take place in 2012.

\section{Court Admissible Evidence}
\label{ch2:legal}

Since the United States leads the way with the implementation of many standards in relation to evidence handling and the court admissibility of evidence, many other countries look to the procedures outlined by the United States in this area when attempting to create their own legal procedures \cite{commons}. As a result of this, much of the information available regarding the admissibility of digital forensic evidence into court cases is specifically tailored to the Unites States, but will influence law makers across the globe. Carrier  \cite{carrier-open} states that in order for evidence to be admissible into a United States legal proceeding, the scientific evidence (a category which digital forensic evidence falls under in the U.S.) must pass the so-called ``Daubert Test" (see section \ref{ch2:daubert} below). The reliability of the evidence is determined by the judge in a pre-trail ``Daubert Hearing". The judge's responsibility in the Daubert Hearing is to determine whether the methodologies and techniques used to identify the evidence was sound, and as a result, whether the evidence is reliable.

\subsection{Daubert Test}
\label{ch2:daubert}


The ``Daubert Test" stems from the United States Supreme Court's ruling in the case of Daubert vs. Merrell Dow Pharmaceuticals (1993) \cite{daubert}. The Daubert process outlines four general categories that are used as guidelines by the judge when assessing the procedure(s) followed when handling the evidence during the acquisition, analysis and reporting phases of the investigation, \cite{carrier-open} and \cite{daubert}:

\begin{enumerate}
\item \emph{Testing} -- Can and has the procedure been tested? Testing of any procedure should include testing of the number of false negatives, e.g., if the tool displays filenames in a given directory, then all file names must be shown. It should also incorporate testing of the number of false positives, e.g. if the tool was designed to capture digital evidence, and it reports that it was successful, then all forensic evidence must be exactly copied to the destination. The U.S. National Institute of Standards and Technology (NIST) have a dedicated group working on Computer Forensic Tool Testing (CFTT) \cite{cftt}.
\item \emph{Error Rate} -- Is there a known error rate of the procedure? For example, accessing data on a disk formatted in a documented file format, e.g., FAT32 or ext2, should have a very low error rate, with the only errors involved being programming errors on behalf of the developer. Acquiring evidence from an officially undocumented file format, e.g., NTFS, may result in unknown file access errors occurring, in addition to the potential programming error rate.
\item \emph{Publication} -- Has the procedure been published and subject to peer review? The main condition for evidence admission under the predecessor to the Daubert Test, the Frye Test, was that the procedure was documented in a public place and undergone a peer review process. This condition has been maintained in the Daubert Test \cite{carrier-open}. In the area of digital forensics, there is only one major peer-reviewed journal, the International Journal of Digital Evidence.
\item \emph{Acceptance} -- Is the procedure generally accepted in the relevant scientific community? For this guideline to be assessed, published guidelines are required. Closed source tools have claimed their acceptance by citing the large number of users they have. The developers of these tools do not cite how many of their users are from the scientific community, or how many have the ability to scientifically assess the tool. However, having a tool with a large user base can only prove acceptance of the tool; it cannot prove the acceptance of the undocumented procedure followed when using the tool.
\end{enumerate}

In 2005, The House of Commons Science Science and Technology Committee in the United Kingdom published a report entitled ``Forensic Science on Trial" \cite{commons}. In this report they outline numerous standards to be used across the field of forensics. As part of this report, the admissibility of expert evidence is discussed. As it stood in the UK when the report was written, the judge of any given case had the role of the ``gate-keeper" for any evidence he would admit into his court. It was determined that judges are not well-placed to determine the scientific validity without input from scientists, especially due to the absence of an agreed protocol for assessment. The main recommendation to come from the report is that the Forensic Science Advisory Council should develop a ``gate-keeping" test for expert evidence, built in partnership with judges, scientists and other key players from the criminal justice system and that it should be built upon the US Daubert Test \cite{commons}.




\section{Summary}
\label{ch2:summary}
This chapter describes some related work to the system described in chapter \ref{ch:arch}. It outlined some of the tools, formats, tests and procedures used for the acquisition and analysis of digital forensic evidence. This chapter also outlined some digital forensic tools and systems developed for aiding digital forensic investigations. Traditionally, in order for a digital forensic investigation to begin, the investigator must physically visit the crime scene and collect any suspect computer equipment. This equipment will then be brought back to the forensic laboratory. Once the computer equipment is in the evidence store, it may then be imaged using one of the forensic tools outlined in \ref{ch2:encase} and \ref{ch2:ftk}. 

\chapter{Design and Architecture}
\label{ch:arch}

\section{Introduction}
\label{ch3:intro}

\begin{figure}
\centering
\includegraphics[width=0.98\textwidth]{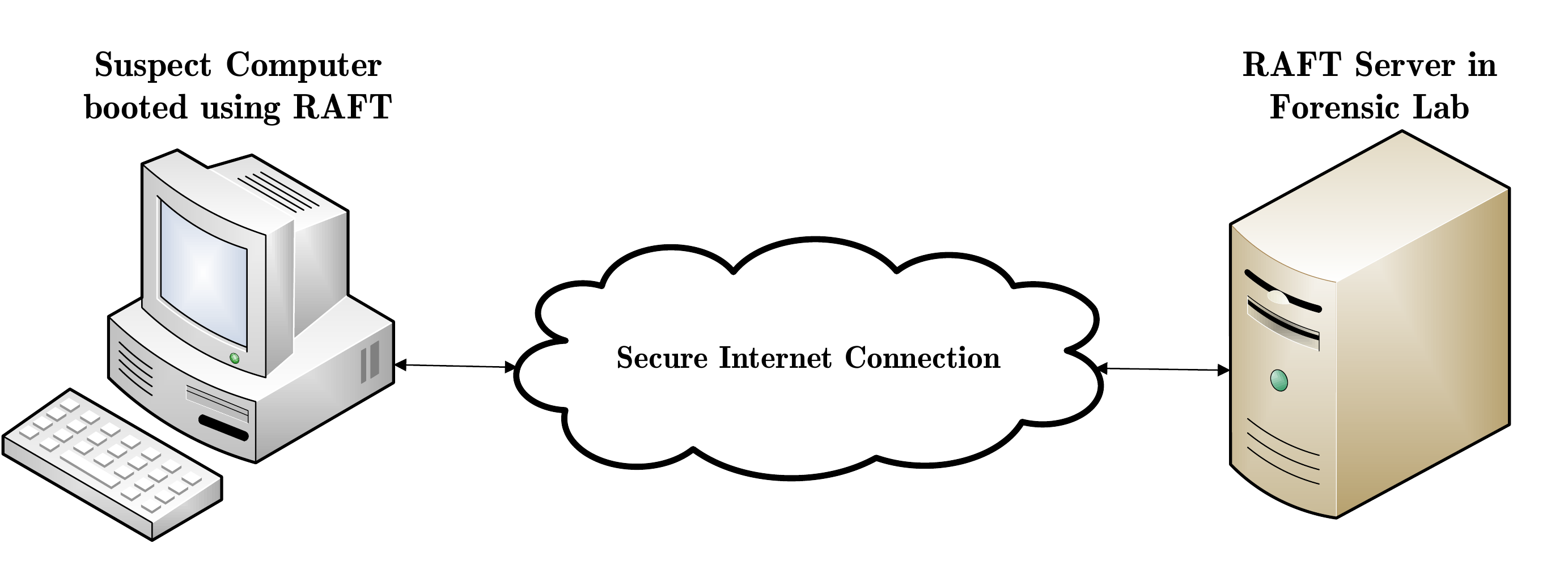}
\caption{Fundamental design of the RAFT system.}
\label{fig:simple}
\end{figure}

The proposed system is concerned with the first step in any digital forensic investigation; the acquisition phase, as outlined in \ref{ch2:evidence}. This system's main goal is to reduce the time taken to acquire the necessary evidence and is called RAFT (Remote Acquisition Forensic Tool). RAFT is a digital forensic hard drive imaging tool designed to boot off a Linux Live CD or USB memory stick and capture a copy of the storage devices on the suspect computer. A brief overview of the RAFT system is depicted in Fig. \ref{fig:simple}. The suspect computer is booted using a customised Linux Live distribution and any hard drive or removable media connected to the computer are able to be securely imaged over an Internet connection directly to the RAFT Server. This system is designed to equip any law enforcement or investigating officers with the ability to easily perform digital evidence acquisition, which would traditionally require the expertise of an on-site forensic investigator. One key objective of the RAFT system is to ensure that the evidence it gathers is court admissible. This is achieved by ensuring that the image taken using RAFT is forensically verified to be identical to the original evidence.

The RAFT system is based on a client/server architecture, as illustrated in Fig. \ref{fig:archnew}. The client side of the RAFT System is designed to be as easy to operate as possible and to require minimal training for the user. In order for the RAFT Client to be used, the suspect computer must be booted from a Live CD or from a USB flash drive. In order for the user to know the procedure involved in booting a computer up, s/he must be trained in the boot selection process and have access to associated documentation. The RAFT server is a multi-threaded server which can accept connections from multiple different RAFT clients simultaneously.

\section{Technical Requirements}
\label{ch3:techrequirements}

When considering the design of any digital evidence acquisition system, such as that described as part of this thesis, it was important to consider some of the technical requirements. These technical requirements include:

\begin{itemize}
\item \emph{Verifiability} -- All evidence collected using the system must be verifiable to the original source. The system must interact with the evidence and suspect system in a forensically sound manner. All evidence interactions must be controllable and reproducible.
\item \emph{Compatibility} -- The system should be capable of working on a variety of computer systems built on many types of different categories of hardware, e.g., servers, workstations, laptops, netbooks etc. The manufacturer of the computer system should have no influence over the performance of the system.
\item \emph{Cost efficiency} -- The cost of implementing the system for any law enforcement agencies or private digital investigators should be as low as possible. Current tools and software can be prohibitively expensive to implement for smaller organisations, e.g., one FRED workstation can cost over \$9000 \cite{fred}.
\item \emph{Usability} -- The client side of the system should not be overly complicated to use. The target user groups, i.e., mainly law enforcement officers, should be considered when designing the user interface. The tool should require minimal training to use.
\item \emph{Scalability} -- The system should be able to scale to any required size. Moore's Law has been commonly used to predict the progression of computer equipment since 1965 \cite{mooreslaw}. Computer processing power, storage capabilities and network speeds will vastly improve in the future. The system should be capable of taking advantage of this predictable advancement.
\item \emph{Multiple User Capability} -- The system should have the ability to be used by multiple users simultaneously. This means that the system must be capable of collecting evidence from numerous geographically separated sources at once.
\item \emph{Extensibility} -- The system must have the ability to be updated with new extensions, such as conforming to a specific digital forensic evidence storage format as outlined in \ref{ch2:storage}.
\end{itemize}

\section{Architecture}
\label{ch3:architecture}

\begin{figure}
\centering
\includegraphics[width=0.98\textwidth]{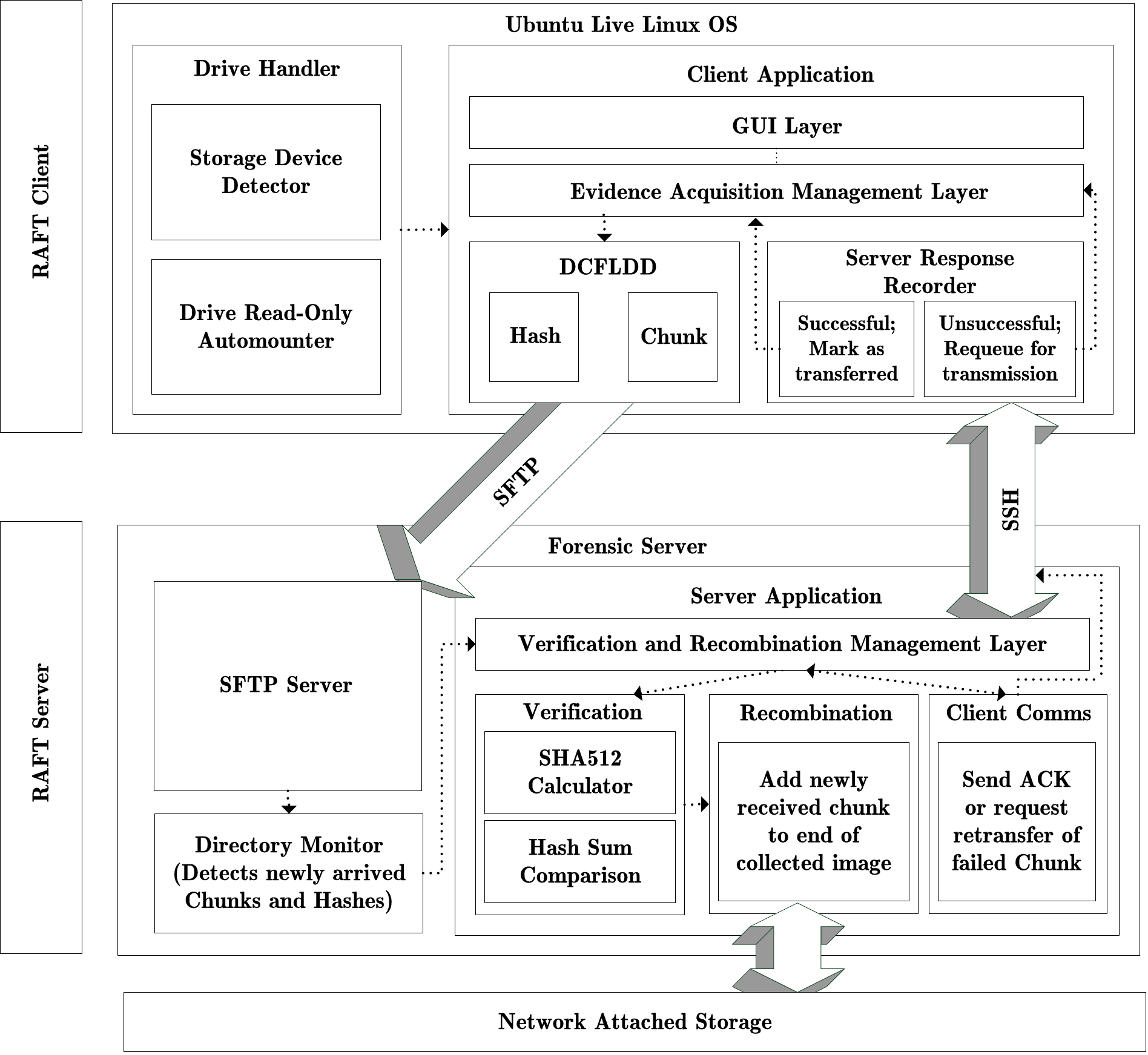}
\caption{RAFT System Architecture}
\label{fig:archnew}
\end{figure}

Figure \ref{fig:archnew} shows the client/server architecture of the RAFT system. The RAFT Client is designed to be booted on the suspect computer. The components incorporated into the RAFT Client's customised Ubuntu Live Linux operating system include the Drive Handler and the RAFT Client application. The Drive Handler detects all storage devices connected to the suspect computer during the boot process and mounts them in the operating system as read-only. The client application deals with the user interaction, evidence acquisition, digital fingerprinting and listening for communications from the RAFT Server over a secure encrypted SSH connection. In this architecture, the evidence acquisition operates over a SFTP connection ensuring the security of the data transfer. The RAFT Server is designed to be located in a forensic laboratory. A SFTP server deals with receiving the collected evidence. The server application deals with the management of the evidence verification and recombination. The final part of the architecture is the Network Attached Storage which is the device which will store all the collected evidence upon successful transmission. 

The components of the RAFT Client and Server are described in greater detail in sections \ref{ch3:client} and \ref{ch3:server}.

\section{RAFT Client}
\label{ch3:client}

\begin{figure}
\centering
\includegraphics[width=0.98\textwidth]{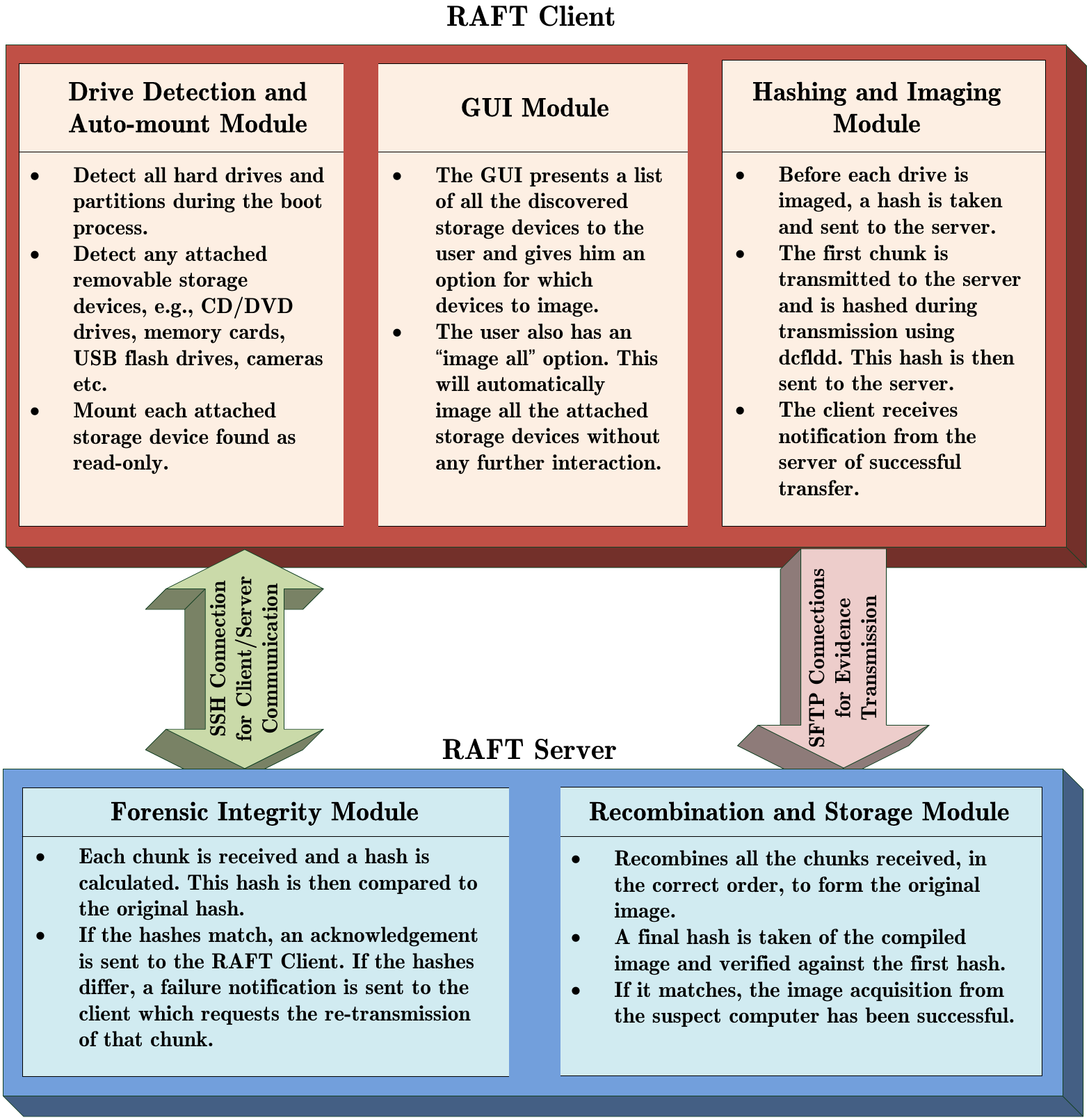}
\caption{Overview of the steps involved, both client and server side, in verifiable image acquisition using RAFT.}
\label{fig:arch}
\end{figure}

The RAFT Client collects forensic evidence as efficiently and quickly as possible and transmits it to the server. It is implemented in such a way as to operate alongside a collection of open source software and tools, e.g., Ubuntu, OpenSSH and DCFLDD. It contains three primary modules which have been integrated into the Ubuntu live operating system; the hard drive and removable media detection and auto-mount module, the graphical user interface module and the hashing and evidence capturing/imaging module, as can be seen in Fig. \ref{fig:arch}.

\subsection{Ubuntu Live CD}
\label{ch3:ubuntu}

The RAFT Client is installed on a customised lightweight copy of the Ubuntu Live Linux distribution \cite{ubuntu}. All unnecessary software and modules should be removed from the operating system to improve performance, e.g., OpenOffice, GIMP etc. Ubuntu is an ideal live operating system to choose for a number of reasons:
\begin{enumerate}
\item The standard Ubuntu install disk comes packaged with a live Linux distribution. This live disk is bootable on any computer, i.e., regardless for which operating system(s) the suspect computer has installed; Windows, *NIX or MAC OS.
\item The compatibility of the live operating system to read numerous different hard drive formats, e.g., FAT, FAT16, FAT32, NTFS, ext, ext2, ext3, HFS, HFS+, etc.
\item The ability for the live distribution to be fully customised removing any unnecessary software, while having the ability to easily include the RAFT Client and associated software (such as dcfldd \cite{dcfldd}, OpenSSH \cite{openssh} and SSH Filesystem \cite{sshfs}).
\item The ability to include a boot up hard drive and removable media auto mounting script which automatically mount all attached drives and storage as read-only.
\end{enumerate}

\subsection{Automatic Drive Mounting Module}
\label{ch3:mounting}

When the Ubuntu Live CD is booted on the suspect computer, all the attached hard drives and removable storage devices currently connected to the host computer are automatically mounted as read-only, e.g., USB flash drives, external hard drives, memory cards, digital cameras, CDs etc. It is imperative that all the connected storage devices are mounted in a read-only state as any accidental writing to the drives could deem the evidence collected as compromised.

\subsection{Hashing and Evidence Transmission Module}
\label{ch3:hashingimaging}

\begin{figure}
\centering
\includegraphics[width=0.8\textwidth]{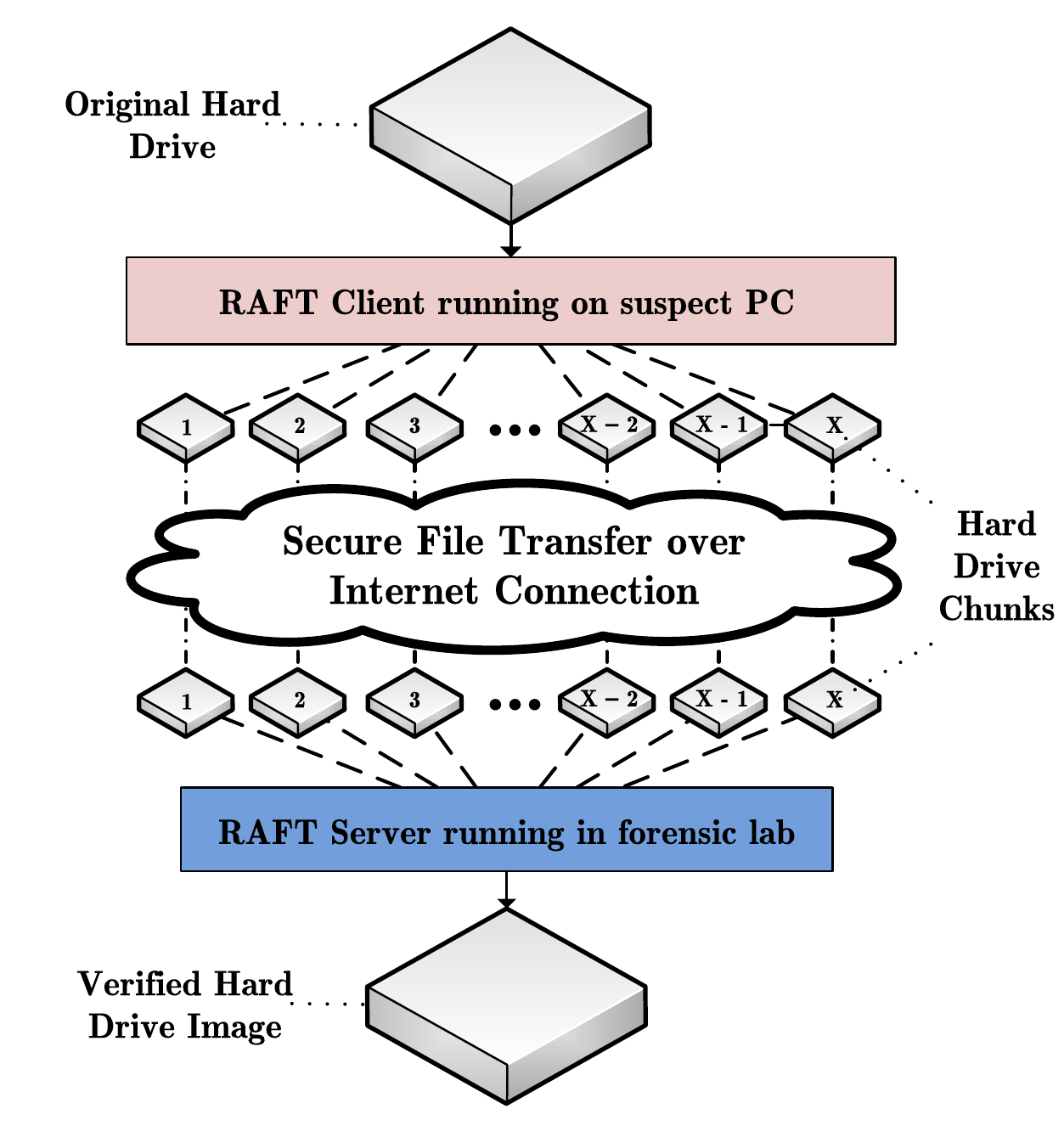}
\caption[Overview of RAFT imaging architecture.]{Overview of RAFT imaging architecture. This diagram shows the regular operation of the RAFT system, i.e., no dropped connections and no failed verifications. X = number of chunks to transfer entire disk image.}
\label{fig:chunk}
\end{figure}

The hashing and evidence transmission module is the segment of the RAFT client responsible for the actual acquisition process. The module instantiates bash shells and executes and monitors the acquisition code. The tool used for the data copying is DCFLDD.

This module is also in communication with the RAFT server. It listens for acknowledgements of successful chunk transfers. Should a chunk fail the verification process, an unsuccessful acknowledgement is received from the server. In this instance, it is necessary to re-add that chunk to the queue for retransmission.

\subsection{Graphical User Interface Module}
\label{ch3:gui}

The graphical user interface is designed to be as intuitive as possible. When the live Ubuntu operating system is finished booting up, the RAFT client is automatically launched. When the application is displayed, there are two options available to the user; ``Automatically Collect All Evidence" and ``Advanced". The first option is self-explanatory, whereas the advanced option displays a list of all detected devices, partitions and removable media. This option enables the user to collect evidence from certain storage devices as a priority over other devices. This option could enable faster analysis of data collected with a high probability of containing incriminating evidence, e.g., in a child abuse investigation, collecting evidence from memory cards or attached cameras may take priority over other data sources.

In a production scenario, the first window presented to the user when launching the RAFT client would be a password entry box. The password for this system would be changed frequently and client side password verification would ensure that the tool could only be used by desirable users. This would eliminate the use of the RAFT client, should a copy of the tool get into the wrong hands. If the password required to state the RAFT client was changed by the server in the middle of a transmission of evidence, it would not affect the transmission as the password is only verified at the start of the RAFT client. This would be a requirement for evidence acquisitions that may take longer than the frequency of server-side password change.

\section{RAFT Server}
\label{ch3:server}

The RAFT Server is a multi-threaded system. When the server is running, it listens for a connection on any of its pre configured ports. When a new client connects to it, it creates a new space on the server where it stores all relevant files. Each new drive or volume imaged from the suspect computer is then stored within that space. The server contains two additional modules for the verification of the collected data chunks and for the recombination of the chunks to form a copy of the original source. The multi-threaded design of the server enables numerous concurrent RAFT Client connections.

\subsection{Forensic Verification Module}
\label{ch3:serververification}

\begin{figure}
\centering
\includegraphics[trim= 10mm 10mm 10mm 10mm, width=0.98\textwidth]{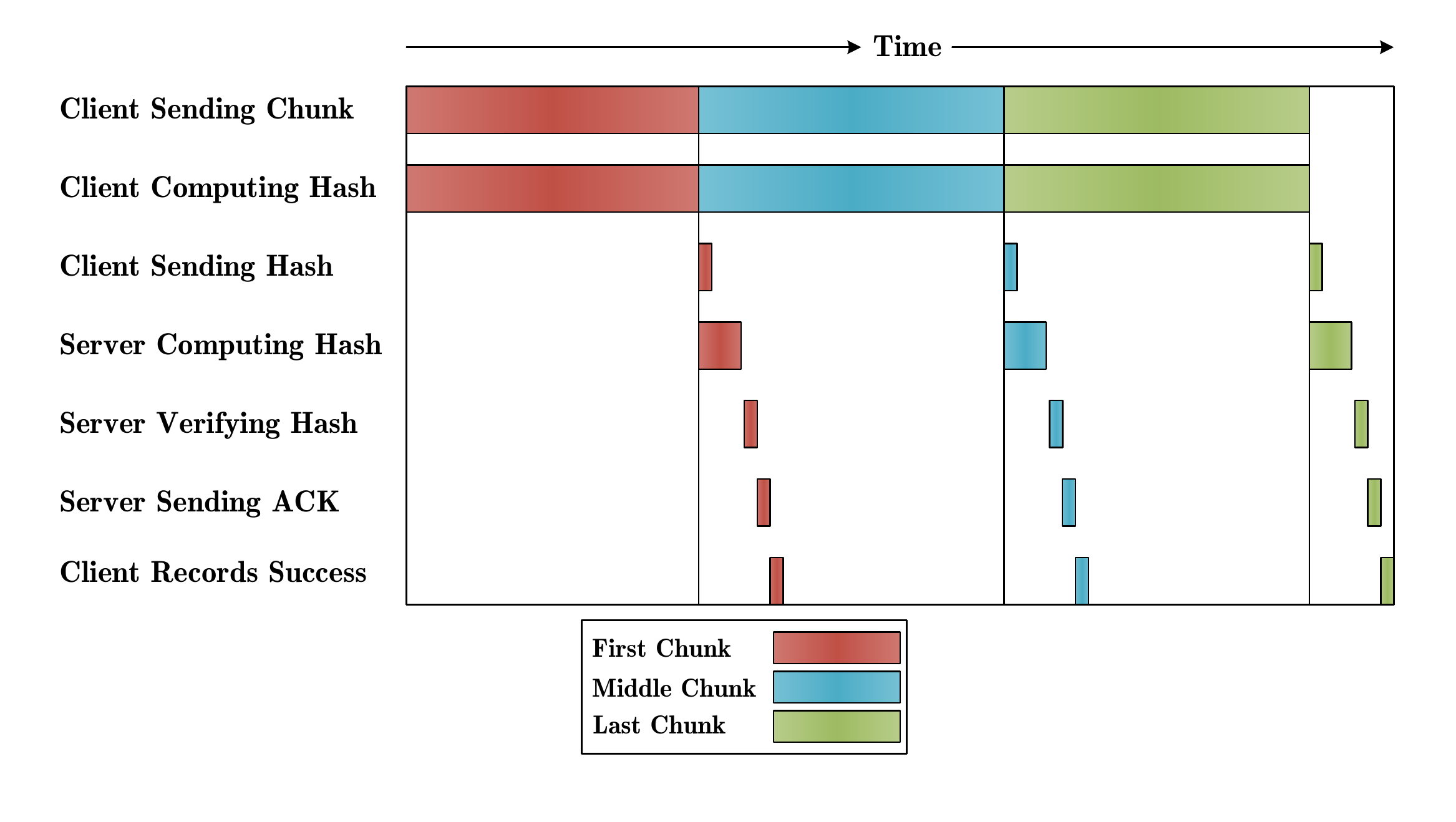}
\caption[Timeline for evidence transmission and verification]{Sample timeline of the RAFT system in operation showing the overall impact of the verification process over the (not to scale).}
\label{fig:timeline}
\end{figure}


Each chunk, when successfully transferred, is hashed on the server and compared against the original hash. If these hashes differ, a failure notification is sent to the RAFT Client which will result in that particular chunk being re-transmitted. A sample sequence of interactions between the client and server for a three chunk acquisition are shown in Fig. \ref{fig:timeline}. As can be seen, the server side verification and recombination of the receiving chunks takes place simultaneously with the receiving of the following chunk.

\subsection{Recombination and Storage}
\label{ch3:recombstorage}
Upon the successful transmission of all the chunks belonging to a particular drive, they are recompiled back into a single file, hashed and verified against the original hash value taken by the RAFT Client before the imaging process commenced. An image acquisition is only successfully complete when this original hash matches which results in a ``completed transfer" notification being sent to the RAFT Client. This recompiled hard drive image is then stored on a network attached storage (NAS) device, similar to the process involved in "traditional" forensic evidence acquisition.



\section{Evidence Handling and Storage}
\label{ch3:handlingstorage}

As with any evidence handling tool, the procedures used to transfer and store the data are important to ensure forensic integrity of the data. The RAFT system is designed to ensure the integrity of the collected evidence at every phase of the acquisition.

\subsection{Evidence Handling}
\label{ch3:handling}

The data collected using the RAFT system is acquired from a read-only mounted hard disk. This ensures that the original source cannot be compromised by any of the operations of the RAFT system. The data is sent directly to the RAFT server in an uncompressed format which ensures that no deleted evidence or any evidence in the slackspace of the drive is lost. Due to the large size of the data transmission, the evidence is sent from the client in chunks and these chunks are then recombined at the server side to produce the complete image of the original disk. As the hard disk image is treated as a binary file during the splitting and recombination processes, the chance of either process compromising the integrity of the data is eliminated. This is proven at the server when the data is recombined and the hash sum of the final image is compared with the untouched hash sum of the original source.

\subsection{Evidence Storage}
\label{ch3:storage}
The ``raw" data storage format outlined in \ref{ch2:raw} was chosen as the storage format for all evidence collected using the RAFT system. This format was chosen for a number of reasons:

\begin{enumerate}
\item The raw format of storing data is the de facto standard for all digital evidence acquisition tools \cite{evidencestandards}. While some tools may have their own proprietary standards, every tool has an option to image the disk in question using the raw format.
\item All evidence analysis tools are capable of reading and analysing the evidence contained in a raw format file.
\item The likelihood of any deleted evidence being destroyed during the imaging process is completely eliminated due to the nature of the format, i.e., each bit of the hard disk is captured including all of the slackspace.
\item Due to the fact that the raw format is an exact bit-by-bit copy of the original evidence, it lends itself well to being split into small chunk sizes, as required by the design of the RAFT system.
\item Acquiring an image using the raw format requires the least amount of processing power client-side. This can be particularly advantageous when collecting evidence from low-powered computers, e.g., older computers, netbooks, etc.
\end{enumerate}

The RAFT system also stores some metadata alongside the evidence collected such as disk information (unique disk identifier, size, partition information, hash sum), number of chunks used to transfer the image and associated hash sums and time stamps of transmission.

\section{Verifying Data Integrity}
\label{ch3:integrity}
The verification of the collected data to the original source is fundamental to the RAFT system. This data integrity is insured in the RAFT system by the implementation of regular hash checking on the data being transferred using SHA-512, a 512-bit secure hashing. Once the RAFT Client is booted and a drive is selected for imaging, the first step is to calculate the hash value for the original drive. During the imaging process, the integrity of each of the chunks being transferred is maintained due to a SHA-512 hash being computed as the chunk is being transmitted. At the server side, once the transmission is complete, a SHA-512 hash is taken on the chunk and verified against the original. If these hashes do not match, i.e. the integrity of that chunk has been compromised in transmission, a failure notification is sent to the client, which queues that chunk up again for transmission.

\subsection{Overhead for Ensuring Data Integrity}
\label{ch3:overhead}

The requirement for any digital forensic evidence capturing tool to ensure integrity is paramount. While one of the primary objectives of RAFT is to verify the integrity of the evidence, it is also important that the additional computational and network overhead is minimised. This is achieved by overlapping the computational tasks with the data transmission.

When the first chunk is transmitted completely to the server, the client immediately starts sending the second chunk. When the server receives the first chunk and its corresponding SHA-512 fingerprint (computed client-side), it then calculates a SHA-512 hash on the chunk received and compares it to the client-side hash. If these hash values match, an acknowledgement is sent to the client to signify a successful transmission. This process is then repeated for the third and all subsequent chunks. Due to the computational/transmission overlap, the additional cost of forensically verifying the evidence captured as part of the RAFT system amounts to the time taken to compute the SHA-512 hash server-side of the last chunk and compare this to the hash value taken client-side.

\section{Resilience against Hacking/Hijacking}
\label{ch3:resilience}


For any digital forensic tool to be considered for use in a real-world law enforcement scenario, the tool must be reliable and resilient against any hacking or hijacking attempts. The RAFT client accomplishes this goal by implementing the following sanctions:

\begin{enumerate}
\item \emph{Encrypted Communication} -- All communications between the client and server are sent over an encrypted SSH connection.
\item \emph{Encrypted File Transfer} -- All data is transmitted using the secure transfer protocol (SFTP), which sends the data through an encrypted tunnel to the server.
\item \emph{Data Verification} -- Each chunk of evidence transmitted is hashed by the client during transmission and subsequently hashed by the server upon receipt. Only when these hashes match is each chunk deemed by the system to be successfully transferred.
\item \emph{Closed Source System} -- The source to the RAFT system is closed source. This improves the system's security by increasing the difficultly for any potential intruders by forcing them to reverse engineer the operational procedures of the system.
\end{enumerate}

In order for the system to be compromised, all of the above sanctions would have to be overcome on-the-fly when the system is in use. While it is currently infeasible to succeed in hacking/hijacking all the above security features, theoretically it may well be possible in the future. The most infeasible part of the above features to hack is to overcome the data verification. The RAFT system uses the SHA-512 hash function, which results in it being particularly resilient against collisions. As of 2009, the only common hash functions greater than 128-bits in length where collisions have been engineered are MD5 and SHA-1. A key point to note about these collisions is that both files were engineered in such a manner to produce the same hash sum, i.e., it is relatively much easier to engineer two files with the same hash when compared to the task of creating a file with a hash that matches an existing file. To compromise the verification process deployed in the RAFT system, the engineered file (with incriminating data removed or with extra data inserted) would have to match the original chunk being transmitted over the encrypted data connection. In the following sub-sections, the difficulty of compromising each step in the acquisition process is discussed.

\subsection{Compromise Client Distribution}
\label{ch3:compromiseclient}

Should a copy of the RAFT client become lost and was acquired by parties interested in cracking the system, they would still need access to the frequently changing server-side password to analyse the operation of the system. Should the attackers reverse engineer the operation of the RAFT client, i.e., determine the data verification process, encryption specifications, etc., this knowledge will still not aid them in cracking the system. The sheer work factor of cracking the SHA-512 algorithm to force a collision results in any such attack being unfeasible.

\subsection{Compromise during Transmission}
\label{ch3:transmission}

In order for the evidence collected by the RAFT system to become compromised during transmission, the following steps must occur on-the-fly during transmission for each chunk:

\begin{enumerate}
\item The encrypted TCP packet stream must be detected and hijacked.
\item The chunk being transmitted must be intercepted.
\item The hash of this chunk must be computed and another chunk with compromised data must be created, with a matching SHA-512 hash sum.
\item This newly created chunk must be transmitted to the server.
\end{enumerate}

Assuming the attacker is aware that the evidence is being transmitted to the RAFT server and is able to detect and identify the encrypted TCP packet stream, he must then successfully hijack the stream. The second step involves intercepting the data stream, decrypting the 128-bit encrypted data (with the exact encryption method not known to the attacker) and building up the chunk being transmitted. This chunk must then be hashed using the SHA-512 hash function (the exact data verification process used in the system will be unknown to the attacker) and another chunk must be engineered with the incriminating evidence removed. This chunk must then be transmitted to the server along the existing TCP stream. Assuming the above steps were feasible, in order for the evidence to be compromised in this manner, a significant amount of system knowledge is required on behalf of the attacker. The combined complexity of each of the above steps results in it being entirely impossible, given the work factor of each of the required steps --- Let alone that it would be possible to occur on-the-fly when the tool is working.

\subsection{Compromise Server}
\label{ch3:compromiseserver}

The RAFT server could be ran on any operating system or hardware configuration. In order for the server to be resilient against attack, only the ports required by the RAFT system should be opened on the machine. A software and a hardware firewall could also be deployed to ensure system security. By limiting the open ports on the server, the chance of the server becoming compromised is reduced significantly. 

\section{Advantages over Traditional Tools}
\label{ch3:advantages}

The RAFT system has a number of advantages over traditional forensic tools. Some of the advantages are valid for using the RAFT system even in a forensic laboratory setting, e.g., using RAFT over a local area network still maintains some advantages over traditional tools. These advantages are outlined in the following subsections:

\subsection{Compatibility}
\label{ch3:compatability}

One obvious advantage of using the RAFT system is that it is irrelevant what configuration the suspect PC has i.e. RAFT is compatible with whatever interface or formatting the suspect hard drive or media might have. Take netbooks as an example: they come in many differing storage configurations, even within the same brand. Some netbooks use regular 2 1/2`` IDE or SATA laptop hard drives whereas some use flash storage. These flash storage devices can be soldered directly to the motherboard, connected via a regular IDE or SATA connection or connected via a mini-PCI/mini-PCIe connection. RAFT has no limitation on what hardware configuration the suspect computer has; the RAFT client is configured to automatically mount and securely image any system configuration.

Should the RAFT client be deployed in a forensic laboratory, the compatibility of the system to collect data from any system is of a significant advantage over the traditional tools. Due to the fact that RAFT is a purely software driven solution, the requirement to have hardware write-blockers at hand for every conceivable storage device connection is eliminated.

\subsection{Cost}
\label{ch3:cost}

The cost involved in running the RAFT system is almost entirely the cost of setting up the RAFT Server. The requirements for the RAFT server would be a high-end computer with the highest speed Internet connection possible (the higher the speed, the less likelihood of running into server-side bottlenecks). It would also be required to have a large amount of available storage, be it local storage or connected network attached storage (NAS). The traditional method of dealing with digital forensic acquisition and analysis involves the storage of hard disk images on a large network storage device. If the additional cost of implementing the RAFT system is compared to the cost involved in purchasing any of the hardware acquisition tools outlined in \ref{ch2:enterprisehardware}, it is clear that it is possible to incur significant savings. These savings could enable a law enforcement branch to afford a fully functional digital evidence acquisition and handling tool. Once the initial outlay is spent in setting up the RAFT Server, the cost for using and re-producing the RAFT Client is minimal. For example, in a law enforcement scenario, the customised RAFT Ubuntu image can be burnt to CD or a bootable USB key and can be created as many times as required, i.e., one of each per police station.

\subsection{Automated Acquisition}
\label{ch3:automated}

This feature of the RAFT system results in users requiring little technological knowledge to operate the client side of the system. Due to the system being designed with ease of use in mind, as outlined in section \ref{ch3:techrequirements}, the adoption of the RAFT system will ultimately result in digital forensic evidence acquisition being possible in more places at once, e.g., in the law enforcement scenario outlined above, each police station would have the capability to image a computer without the need to have a digital forensic specialist.

\subsection{Speed}
\label{ch3:speed}

The RAFT system enables digital evidence to be captured by any law enforcement officer in more spaces simultaneously. In the current digital evidence acquisition model used by law enforcement, when a regular police officer is at a crime scene and identifies one or more computers as potential evidence sources, he must request a digital forensic investigator to travel to the crime scene and collect the evidence. If multiple cases require evidence collection at the same time, there can be a significant amount of wasted investigation time. While each individual image acquisition can take some time, multiple acquisitions can take place simultaneously and results in an overall decrease in the time taken for multiple computers to be imaged.

\section{Overcoming Limitations}
\label{ch3:limitations}

While the RAFT system has several advantages over the traditional approach, such as those outlined above, there are also some potential limitations as outlined in the following sub-sections. For each of the potential limitation described, a potential solution is provided.

\subsection{Firewalls}
\label{ch3:firewalls}

The RAFT Client has to have the ability to communicate to the server, for the transmission of the evidence. One obvious potential limitation of the system is that a hardware firewall may be filtering the suspect computer's Internet connection i.e. banning specific port ranges etc. This could potentially render the RAFT Client inoperable. One solution to this is to employ the use of a USB mobile broadband connection, connected to the suspect computer. Current 3G wireless broadband networks are capable of upstream speeds of up to 10 Mbps, with plans for 3G LTE (Long-Term Evolution) to increase the upstream speeds to over 50 Mbps \cite{3g}. These potential upload speed are set to improve even further when 4G mobile broadband networks become mainstream in the coming years. 4G networks will be capable of upload speeds of over 100 Mbps \cite{4g}. If a software firewall is installed on the suspect computer, it is only installed to monitor the network traffic from the operating system on the suspect computer. This will not affect the operation of the RAFT system, as the suspect computer will then be running the customised Ubuntu Linux operating system.

\subsection{Transfer Speed}
\label{ch3:transferspeed}

The time taken to take an image of a hard drive over the Internet will take longer than the time required if the investigator had physical access to specialised forensic hardware in a forensic laboratory. Where RAFT can improve on this time required for traditional hard drive image acquisition is if the time wasted by the investigation in travelling, transportation and storage of the suspect computer is taken into consideration. While high-speed broadband Internet access is becoming more and more common place on both residential and commercial levels, it would be unrealistic to assume that every suspect computer would have an Internet connection with a favourable upload speed, i.e., many asymmetric broadband connections are significantly weighted towards download speeds. This limitation could again be overcome through the use of a mobile Internet connection.

\subsection{Potential LiveCD Incompatibility}
\label{ch3:liveCD}

It is possible that some suspect computers have a non-functional CD drive. For that matter, some modern portable computers do not feature a CD/DVD drive, e.g., most netbooks and small laptops. This issue can be overcome as the customised Ubuntu operating system containing RAFT can also be configured to boot off of a USB memory key. In the case of netbooks, each version available on the market today is capable of booting from a USB flash drive or USB CD drive as these are the only methods of loading the operating system onto the computer.

\subsection{Live System}
\label{ch3:livesystem}

Forensic investigators are increasingly concerned with the analysis of live systems, e.g. collection of evidence of current processes, memory and other state information. In its current from, the RAFT system is unable to collect evidence from a live system. However, a modified version of the RAFT client could be created to run on a live system. The downside of executing RAFT on a live system is that there will be an unavoidable, yet predictable, change of state of the live system. This will be one of the challenges in building such systems, as any modification to the system, e.g., plugging in a UCD key, inserting a CD or executing an application, will modify its state, e.g., modify the registry, log files, volatile memory etc.

\subsection{Boot Passwords}
\label{ch3:bootpasswords}
The CMOS (Complementary Metal-Oxide Semiconductor) on modern motherboards is responsible for semi-permanently storing system information such as the clock, memory amount, hard drive information, BIOS boot password along with other system configuration settings. The ROM is powered by a small battery when the system is powered down. If the suspect system is configured with a boot password, before the user has the opportunity to boot up the RAFT Client, s/he must insert their password. In the quite likely event that this password is unknown, there are three options available to the user:

\begin{enumerate}
\item Reference documentation -- All motherboard manufacturers incorporate a backdoor CMOS password into the BIOS. For the user to determine the correct password it will be necessary to reference documentation to retrieve this password. A sample list of common BIOS manufacturer's and their associated backdoor passwords can be see in table 3.~\ref{tab:passwords} \cite{wang}, \cite{scanlon}.

\begin{table}
\label{tab:passwords}
\begin{center}
\begin{tabular}{| l | p{9cm} |}
\hline
Manufacturer        & Commonly Used Passwords \\
\hline
\hline
AWARD           & 01322222, 589589, 589721, 595595, 598598 , ALFAROME, ALLY, ALLy, aLLY, aLLy, aPAf, award, AWARD PW, AWARD SW, $AWARD?SW$, $AWARD\_PW$, $AWARD\_SW$, AWKWARD, awkward, BIOSTAR, CONCAT, CONDO, Condo, condo, d8on, djonet, HLT, J256, J262, j262, j322, j332, J64, KDD, LKWPETER, Lkwpeter, PINT, pint, SER, $SKY\_FOX$, SYXZ, syxz, TTPTHA, ZAAAADA, ZAAADA, ZBAAACA, ZJAAADC \\
\hline
AMI                 & AMI, AAAMMMIII, BIOS, PASSWORD, HEWITT RAND, $AMI?SW$, $AMI\_SW$, LKWPETER, A.M.I., CONDO \\
\hline
PHOENIX             & BIOS, CMOS, phoenix, PHOENIX, Phoenix \\
\hline
\end{tabular}
\end{center}
\caption{Backdoor BIOS passwords for common motherboard manufacturers}
\end{table}

While such a list may seem cumbersome if each password was attempted sequentially, a database of computers and motherboards and their corresponding backdoor passwords could be created to quickly facilitate access to the correct password.

\item Hardware Password Bypass -- Some motherboards contain a jumper  on the chip which, when removed, lets the user bypass any passwords contained in the CMOS. This physical backdoor was incorporated into the design of the motherboard as to eliminate the possibility of a customer getting locked out of their system.

\item CMOS Reset -- When the CMOS battery is removed from the motherboard for a short period of time, e.g. less than ten minutes, the CMOS will reset to its original state, with the factory settings for the boot password. In most motherboards, the default factory configuration is not for this password to be activated on boot.

\end{enumerate}

Passwords further in the regular boot process of the suspect system, e.g. a Windows login password, will have no effect on the operation of the RAFT system as it will be the customised Live Ubuntu operating system which is booted immediately after the BIOS.

\section{Adoption for Different Users}
\label{ch3:adaption}

To configure the system for different users, i.e., different law enforcement departments, different private investigation companies, there are a number of items in the system that will need to be configured:

\begin{enumerate}
\item \emph{Hard Coded Hostnames} -- Each RAFT client is built and configured for a particular user. The live operating system needs to be modified with hard-coded hostnames. In a production system, this would also have to be modified with a SSH key for communication with the server.
\item \emph{Hard Drive Formats} -- While the standard RAFT client is capable of reading from numerous hard drive formats, as outlined in section \ref{ch3:ubuntu}, the tool may be required to be configured to take evidence from a uncommon proprietary format.
\item \emph{Hardware Drivers} -- The drivers built in to the Ubuntu Linux are compatible with the majority of hardware configurations available. Should the RAFT client be expected to capture evidence from an uncommon hardware device, the driver for this device may need to be included in the live operating system.
\item \emph{Any Further Customisation} -- This would include any customer specific requirements to customise the system to be compatible with their current infrastructure. This could include adjusting the evidence storage format to comply with any of the formats outlined in section \ref{ch2:storage}.
\end{enumerate}

\section{Summary}
\label{ch3:summary}
The security of the RAFT system is paramount, as with any evidence handling device. By limiting the port ranges that are left open to access from the Internet, the chance of an attack on the server is also limited. The server would also be protected by a software and a hardware firewall; only allowing encrypted SSH and SFTP traffic through to the server.

This chapter also outlined a number of the advantages the RAFT system would have over traditional tools. While no forensic tool (hardware or software) on the market today is capable of dealing with every possible scenario, the RAFT system is capable of overcoming many of the limitations a remote evidence acquisition tool might face. The combined advantage of the outlined points results in the forensic investigator being able to spend more time in the laboratory analysing the evidence collected, as opposed to time wasted performing menial tasks, e.g., travelling to crime scenes. Using RAFT in combination with more intelligent forensic analysis tools, e.g., a distributed digital forensic system \cite{gao}, \cite{richard}, \cite{roussev}, \cite{kulesh}, the investigator will be better armed to deal with the ever increasing amount of digital forensic cases.

The only comparable existing solution to the system outlined above is that of the Forensic Recovery of Evidence Device Diminutive Interrogation Equipment (FREDDIE), as outlined in section \ref{ch2:freddie} above. This product is comparable with the RAFT system, in so far as they both enable on-site evidence acquisition. A notable point to differentiate the two tools is that the cost of purchasing each FREDDIE device starts at \$8,000 \cite{freddie} (price correct as of August 2009), whereas the cost of creating an additional copy of the RAFT client, is purely the cost of burning a CD or making a copy of the bootable USB flash drive. The use of a FREDDIE also maintains the existing requirement for the digital forensic investigator to physically visit the crime scene. With RAFT, any law enforcement officer has the immediate on-site capability of collecting digital evidence from any machine.

\chapter{Experimentation and Results}
\label{ch:results}

\section{Introduction}
\label{ch4:introduction}

This chapter discusses the results of testing the prototype of the RAFT system. The RAFT client was implemented through the development of an application and installing on a customised lightweight Ubuntu Linux LiveCD.
This chapter also includes a performance evaluation of the RAFT system. As part of the testing process, numerous ``real-world'' scenarios were tested and for the purpose of this thesis, we will discuss two of these scenarios.

\section{Viability Testing}
\label{ch4:viability}

To test the viability of the RAFT system, each component of the client and server needed to be prototyped and tested individually. The RAFT client has two main components that require testing; the hard disk and storage device detection and write-blocked mounting component and the evidence acquisition and hashing component. To test the viability of the server side of the RAFT system, it was required to test the hashing and verification component and the recombination and final verification component. The communication between the system also needed to be tested to decide upon which protocol to use. 

In order for any digital forensic tool to be considered for use in a law enforcement scenario, the evidence it collects must be proven to be unmodified and the tool must be proven to be reliable with reproducible results. In order to prove this, the tool was tested to verify if the evidence collected was identical to the original evidence source.

The following subsections outline the results from the viability studies conducted.

\subsection{Hash Function Experiments}
\label{ch4:verification}

In order to choose the most secure hash function, i.e., the most resilient to collisions, while still maintaining a relatively low time overhead to the acquisition process, a number of hash functions were performance tested. These include some of the more popular hash functions, as can be seen in Table \ref{ch2:hashchoice}. In the tests outlined below the hash functions deemed the most secure (SHA-224, SHA-256, SHA-384 and SHA-512) are performance tested uses various input sizes to attempt to analyse the performance of each function according to the differing input message sizes. The hashing times of the MD5 and SHA-1 functions are also included for reference, although these hash functions have been compromised to some extent resulting in engineered collisions, as outlined in more detail in section \ref{ch2:collisionresistance} above. To create the test files for hashing, ``zeroed-out" files of the exact specified size were created using the *NIX ``dd" command. For example:

\begin{verbatim}
dd if=/dev/zero of=1GBFile bs=1073741824 count=1
\end{verbatim}

When executed on a *NIX system, this command creates a file called ``1GBFile" with the input being a null file and the output file being of the specified bytesize (1,073,741,824 bytes = 1GB).

\begin{figure}
\centering
\includegraphics[trim= 30mm 30mm 30mm 30mm, width=0.98\textwidth]{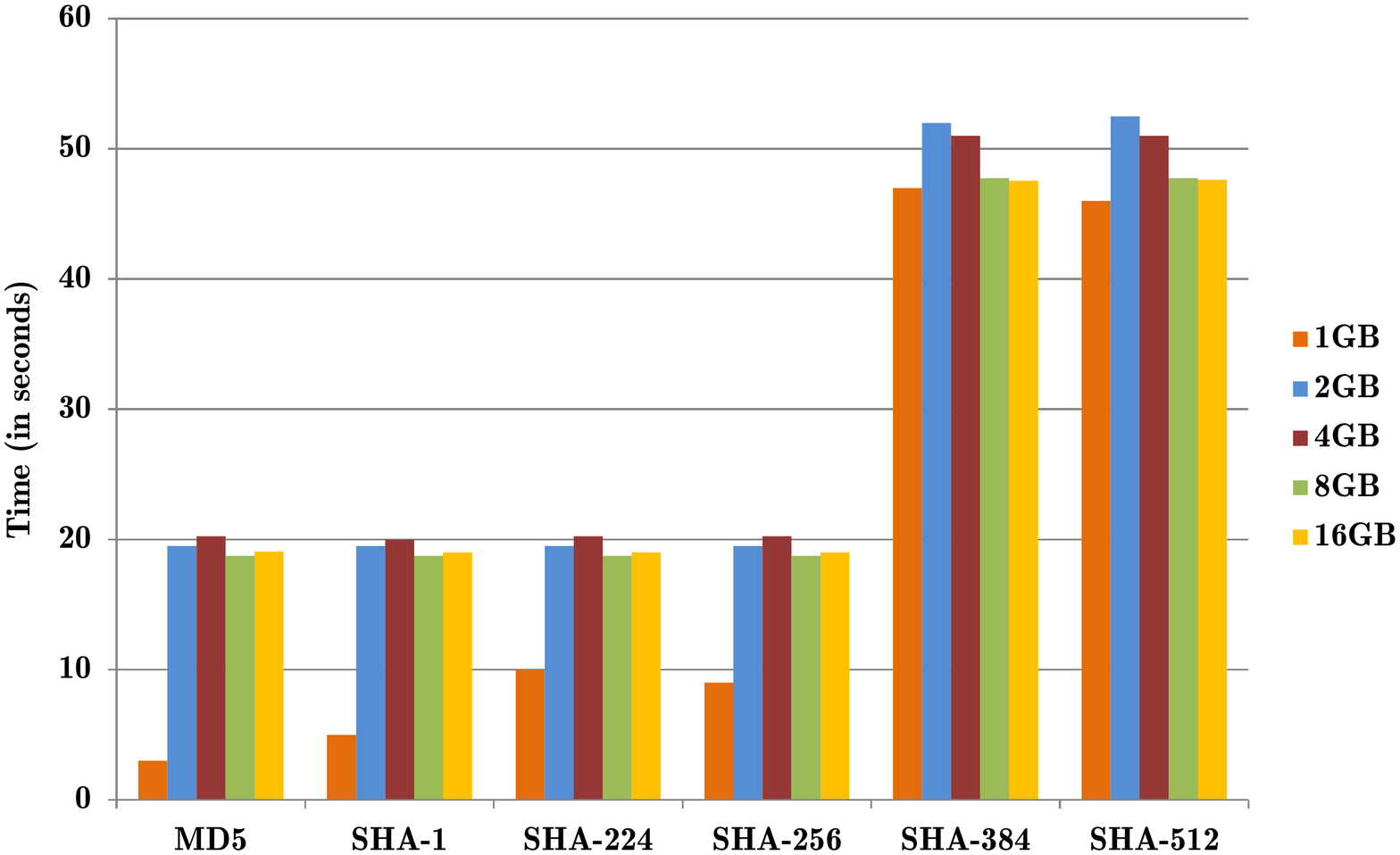}
\caption[Comparison of server side hashing times]{Comparison of server side hashing times for differing image sizes (averaged to a per-gigabyte value).}
\label{fig:averaged}
\end{figure}

Figure \ref{fig:averaged} represents a bar chart of the results obtained for testing the various indicated hash functions. This test was conducted  with a Dell Optiplex 745 with a 2.66Ghz Intel Core 2 Duo processor, 2GB 667Mhz memory and a 250GB 3.5`` 7200rpm hard drive representing the server side of the RAFT system. The Figure shows the results, averaged out to a ``per-gigabyte" value, of the creation time for hash sums for 1GB, 2GB, 4GB, 8GB and 16GB files using each of the hash functions. As can be seen from the chart, the average time taken for hashing using the functions with 256-bit and below internal state sizes is approximately linear, i.e., MD5, SHA-1, SHA-224 and SHA-256. The exception to this is the time taken for these algorithms to produce hash sums for the 1GB input message. This is due to the algorithms having efficiency for hashing smaller files. The average time required for these hash functions to produce a hash in these tests was 19.36 seconds with a variance of $\pm$ 0.41\% per gigabyte. The time required for the algorithms with internal state sizes of greater than 256-bits was consistently found to be over double that of the other functions outlined above. The average time for SHA-384 and SHA-512 was found to be 49.01 seconds with a variance of $\pm$2.89\% per gigabyte.

\begin{figure}
\centering
\includegraphics[trim= 30mm 30mm 30mm 30mm, width=0.85\textwidth]{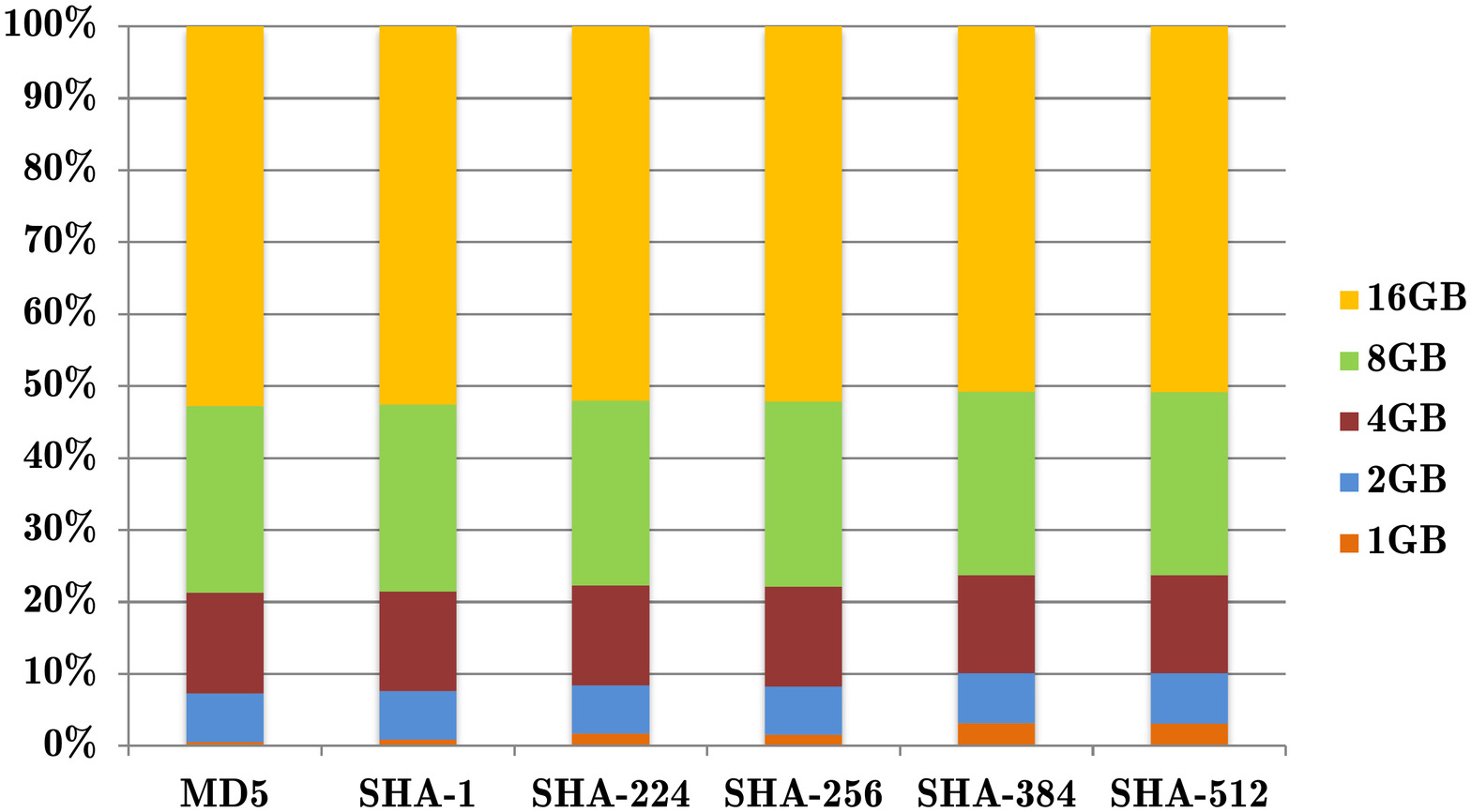}
\caption[Bar chart displaying linear hash time increases]{Bar chart showing the almost linear time increases for the differing input sizes. 100\% represents the time taken to produce the hash sum for each hash function individually for the largest file tested (16GB).}
\label{fig:linear}
\end{figure}

The time required during testing to produce the server side hash sums for various file sizes is outlined in Fig. \ref{fig:linear}. As can be seen, this time is almost on a linear scale, i.e., the time taken for each hash sum to be produced approximately doubles when the input message size doubles. During the data transmission phase of the RAFT system, these server side values are calculated on the server upon receipt of each chunk. The hash value for chunk X is computed when the client is transmitting chunk X+1. This method of simultaneous data transmission and verification results in the overhead created due to the verification process being as minimally impacting as possible.

\begin{table}
\label{ch4:400gbhash}
\begin{center}
\begin{tabular}{| p{5cm} | p{8.1cm} |}
\hline
File Variation      & SHA-512 Hash \\
\hline
One byte file containing the single character ``0'' &  \texttt{F8C7E45008E5EAED1573C6E41FA7F172
8CC7A193C17759D858F0F4757A862A97
77A2825E9CE7ACD8272A69DB079AE8E4
F563996C1EA7C28AFF34EA373D98F9AE}\\
\hline
One byte file containing the single character ``1''  &  \texttt{3C040891330C17E1F66951EE2DFA47E0
C05DCA805CA68B48433AAB582D79272A
6C1C0A51B3BABFCC055AF823729D8EE1
C314221BCC61476C9133F0378BBA40CB}\\
\hline
\end{tabular}
\end{center}
\caption[Avalanche effect of SHA-512]{Two SHA-512 hash sums of a 400GB hard drive. One file was edited between the tests by replacing one byte, i.e., changing a ``0'' to a ``1''.}
\end{table}

Table \ref{ch4:400gbhash} shows the impact of the avalanche effect, as outlined in \ref{ch2:avalanche}, to the SHA-512 hash function. For this test a 400GB hard drive was hashed twice with a minor one byte modification between tests. As can be seen, the resultant hash sum is completely different. The chance of a collision occurring when using the SHA-512 is approximately 1 in 1.34 x $10^{154}$. If this probability of a collision is compared with other commonly court admissible forms of uniquely identifying evidence, it is found to be vastly superior. For example, the chance of a collision occurring in Human DNA evidence is approximately 1 in 1 x $10^{41}$ \cite{dna} and the chance of a collision occurring in commonly accepted human fingerprint evidence is approximately 1 in 5.86 x $10^{7}$ \cite{fingerprint}.

\subsection{File Transfer Protocols}
\label{ch4:ftp}

During the testing process for the various possible file transfer protocols that could be implemented for the RAFT system, it was found that regardless of which protocol that was used, the transfer time for any sized files was within a $\pm$0.02\% deviation. The file transfer protocols tested included:

\begin{enumerate}
\item \emph{File Transfer Protocol (FTP)} -- This protocol sends the data through an unencrypted connection to the server. While sending the data through an unencrypted connection is not best practice for sending forensic evidence, this protocol was included in the testing phase as a comparison for the other secure protocols.
\item \emph{Trivial File Transfer Protocol (TFTP)} -- This is a very lightweight unencrypted file transfer protocol. Its main advantage is that due to its lightweight and simple nature, it requires very little memory to run. 
-\item \emph{FTP over SSH (FTPS)} -- This protocol is also known as FTP-SSL. It was the first secured, encrypted file transfer protocol created with the aid of the Secure Socket Layer (SSL) wrapper. The authentication is dealt with through the use of certificates. While FTPS performed well in testing, it does not have the ability to resume transfers and for this reason it was deemed unsuitable for the purposes of the RAFT system.
\item \emph{Secure Copy (SCP)} -- The Secure Copy Protocol is built on the BSD "rcp" protocol. The security of this protocol is handled by the underlying SSH protocol. This protocol was generally found to be marginally faster than the other protocols tested due to a more efficient transfer algorithm. However, this extra efficiency comes at the expense of reliability of the transfer, i.e., SCP does not require confirmations of successful data transmission nor can it continue interrupted transfers. Weighing the minor time difference against the reliability, this protocol was deemed unsuitable for the use in the RAFT system.
\item \emph{SSH File Transfer Protocol (SFTP)} -- SFTP is sometimes referred to as Secure FTP or SSL-FTP. This protocol handles security by being built on top of the SSH protocol. SFTP has the ability to resume interrupted data transfers without needing to resend the already transmitted data. While not being the fastest protocol overall, the SFTP protocol was just marginally slower than SCP while having a higher degree of reliability and the ability to resume interrupted transfers. For these reasons, SFTP was decided upon as the ideal protocol for the RAFT system.

\end{enumerate}


\subsection{Recombination}
\label{ch4:recombination}

The recombination of the chunks received is handled using the *NIX ``cat" command. This command is used to append each received chunk, in the correct order, onto the recombined image. In testing, the time taken by the server to append each chunk onto the end of the recombined image was less than one second.

A sample usage of the cat command used to append each chunk received to the end of the recombined image is shown below:

\begin{verbatim}
cat chunkX >> combinedimage
\end{verbatim}

\section{``Real World" Experiments}
\label{ch4:experiment}
To evaluate the performance of the RAFT system, numerous scenarios were tested. For the purpose of this thesis, we will discuss two of these scenarios showing the typical usage of the RAFT system in residential and enterprise evidence gathering.

\subsection{Residential Experiment}
\label{ch4:residential}

\begin{figure}
\centering
\includegraphics[trim= 10mm 10mm 10mm 10mm, width=0.98\textwidth]{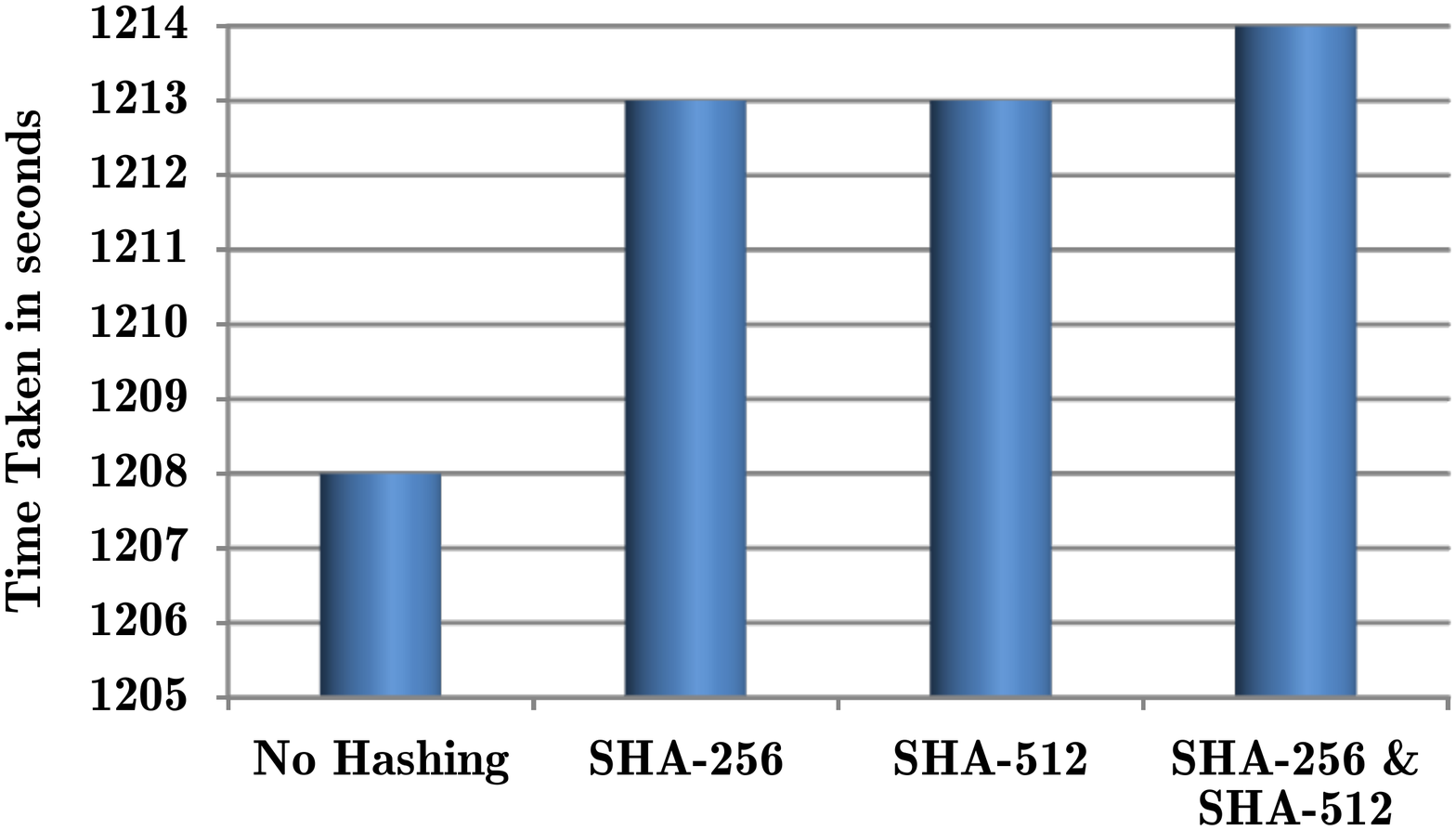}
\caption[Evidence collection graph from an average residential Internet connection.]{Comparison of imaging per gigabyte times from a residential broadband connection with a 8.14Mbps upload speed as tested using speedtest.net.}
\label{fig:homehash}
\end{figure}

The first scenario involved imaging a suspect computer from a residential broadband connection, with relatively low bandwidth speeds. The suspect computer in this scenario was a Dell XPS M1330 laptop with a 2.5Ghz Intel Core 2 Duo processor, 4GB 667Mhz memory and a 320GB 2.5'' 7200rpm hard drive. The broadband connection had a download speed of 23.82Mbps and an upload speed of 8.14Mbps (connection speed tested using Speedtest \cite{speedtest}). The result from these tests showed that the average time to acquire a 320GB hard drive image was approximately 20 minutes per gigabyte, as can be seen in Fig. \ref{fig:homehash}.

\subsection{Enterprise Experiment}
\label{ch4:enterprise}

\begin{figure}
\centering
\includegraphics[trim= 10mm 10mm 10mm 10mm, width=0.98\textwidth]{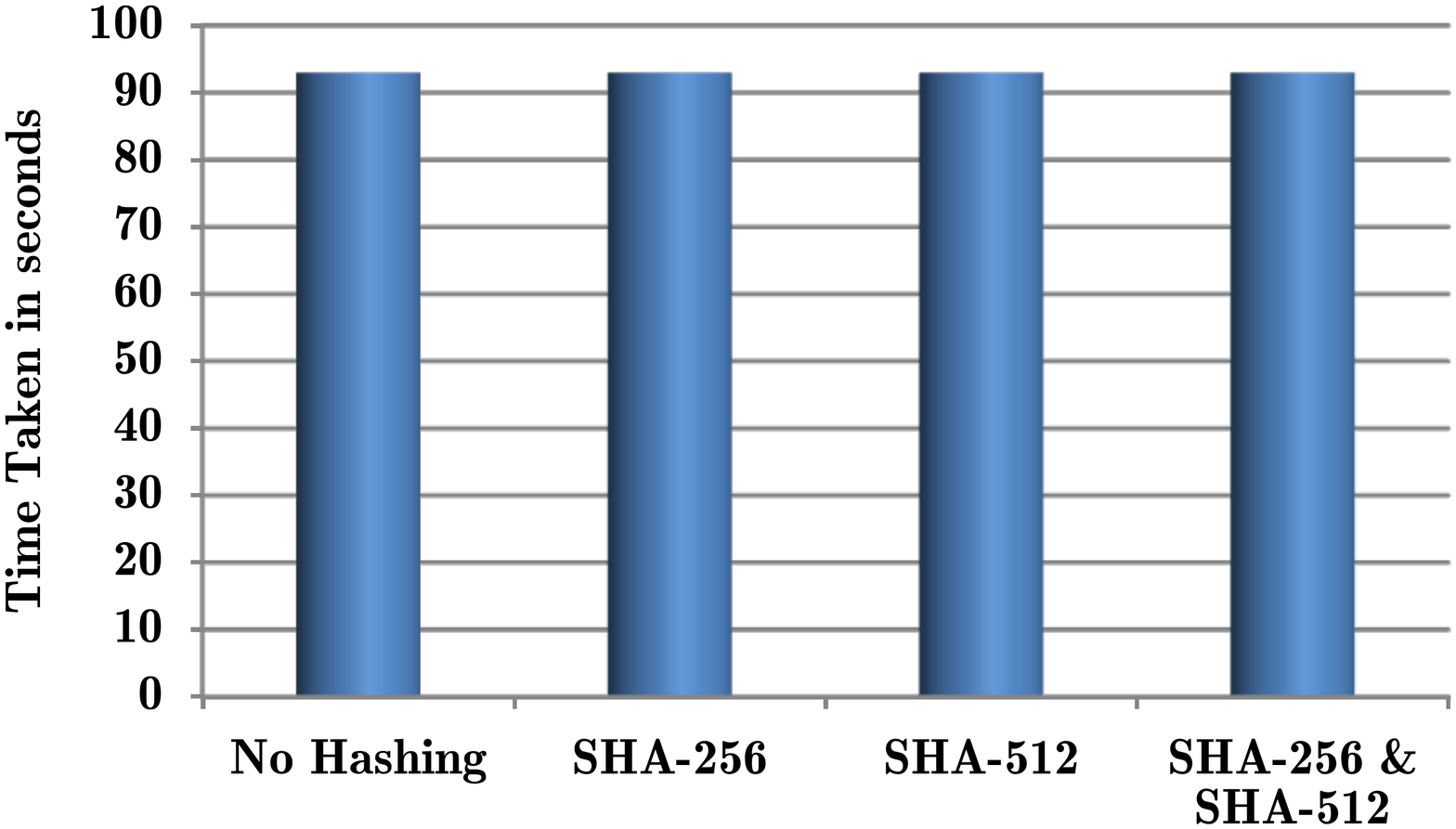}
\caption[Evidence collection graph from a high-speed enterprise Internet connection.]{Comparison of imaging times per gigabyte from a high-speed Internet connection with a 83.26Mbps upload speed as tested using speedtest.net.}
\label{fig:ucdhash}
\end{figure}

The second scenario incorporated imaging a target computer with a very high-speed Internet connection with a 87.62Mbps downlink and a 83.26Mbps upload streams (connection speed tested using Speedtest \cite{speedtest}). The suspect computer in this scenario was a Dell Optiplex 745 with a 2.66Ghz Intel Core 2 Duo processor, 2GB 667Mhz memory and a 250GB 3.5`` 7200rpm hard drive. As can be seen in Fig. \ref{fig:ucdhash}, the average time required per gigabyte was 92 seconds.

\section{Results}
\label{ch4:results}
During the testing of the RAFT system, the performance of the imaging process tended to be linear. As a result, all of the results discussed below have been normalised to reflect the average performance for one gigabyte. The ``dcfldd'' tool used in the RAFT system has the ability to compute the hash values at the same time as transmitting the chunk. The four values displayed in Figures \ref{fig:ucdhash} and \ref{fig:homehash} show the impact of the various hashing options on the overall performance.

\begin{figure}
\centering
\includegraphics[trim= 10mm 10mm 10mm 10mm, width=0.98\textwidth]{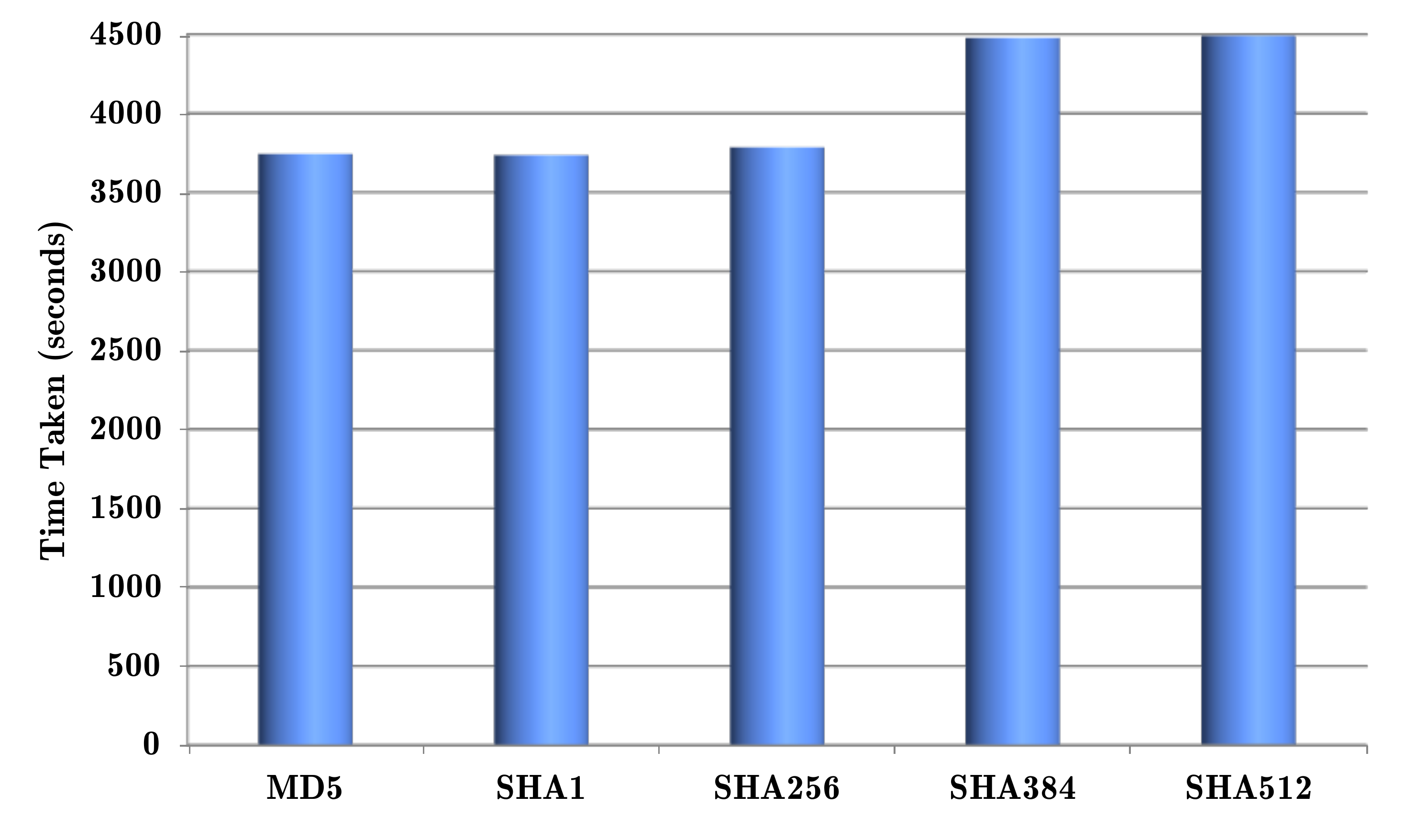}
\caption{Performance comparison of hashing algorithms according to time.}
\label{fig:hashgraph}
\end{figure}

One requirement of the performance evaluation of the RAFT system was to quantify the overhead added through the secure hashing of each chunk. It was found that the cost for the hashing of each chunk averaged at 5.3 seconds per gigabyte (or a 0.41\% increase in the time taken) as can be seen in Fig. \ref{fig:homehash}.

\begin{figure}
\centering
\includegraphics[trim= 30mm 30mm 30mm 30mm, width=0.98\textwidth]{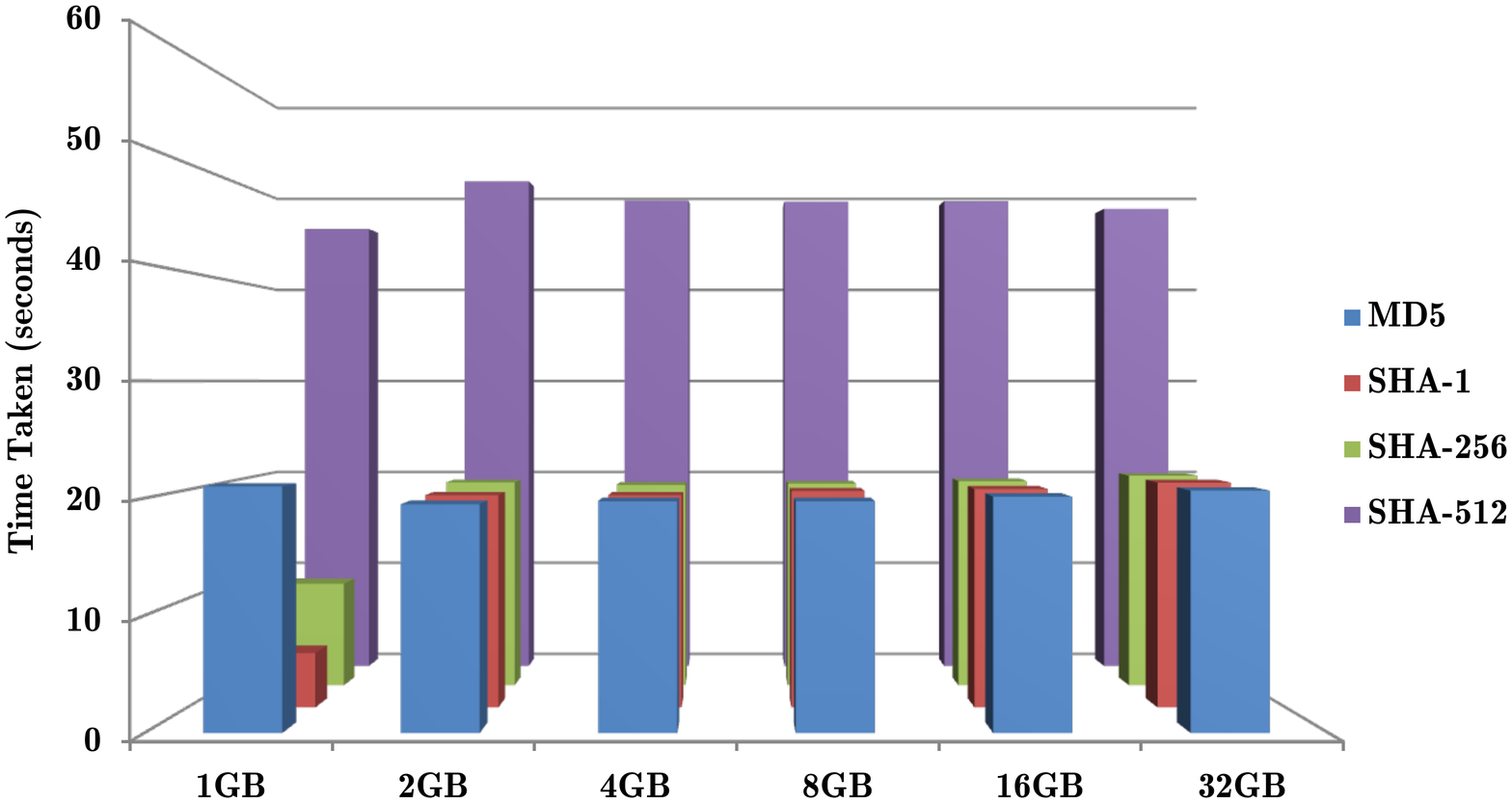}
\caption{Server side comparison of selected common hashing algorithms on various file sizes (normalised to a per-gigabyte value).}
\label{fig:serverhash}
\end{figure}

The time taken for the server to verify each of the hard drive chunks received is approximately 20 seconds per gigabyte using the SHA-256 hashing algorithm, as can be seen in Fig. \ref{fig:serverhash}. This figure also displayed the server-side hashing times as a comparison of three other common hashing algorithms. As can be seen, there is no additional overhead involved in choosing to use the 256-bit algorithm (SHA-256) as opposed to the 128-bit algorithm (MD5) and the 160-bit algorithm (SHA-1). However, there is a substantial cost of over double the computational time in using the more secure 512-bit algorithm (SHA-512). The extra time required to compute the SHA-512 hash will only impact the overall imaging time twice; once on the RAFT Client before imaging begins, and again when the image is completed.

Assuming no dropped image chunks during transmission, the total time (T) required for a remote image acquisition of a hard drive size (H), chunk size (C), with a client-side broadband upload speed (B) and a server side hashing and verification speed (V) can be summarised by the following formula:

\begin{equation}
T = \dfrac{H}{B} + \dfrac{C}{V}
\end{equation}


\section{Evidence Capture Overview}
\label{ch4:overall}

The total time taken for a complete acquisition from any suspected computer can be split into four influencing factors:

\begin{enumerate}
\item \emph{Total transmission time} -- This is the total time required to transfer the entire disk image, chunk by chunk, from the target computer to RAFT server. This time value will need to incorporate the retransmission time for any dropped connections or any chunks which fail the server side verification process. It is notable that during testing of the prototype, neither of these conditions that would require re-transmission were encountered.
\item \emph{Time to produce hash value for the final chunk} -- Only the hashing of the final chunk needs to be taken into consideration when calculating the total time required to take a complete image of the suspect as the hash value for every other chunk is calculated and verified on the fly during the transmission its proceeding chunk.
\item \emph{Recombination of the chunks} -- This is the time it takes for the server to recombine all the received chunks into a single exact replicated file of the original evidence.
\end{enumerate}

\begin{figure}
\centering
\includegraphics[trim= 25mm 20mm 0mm 0mm, width=0.75\textwidth]{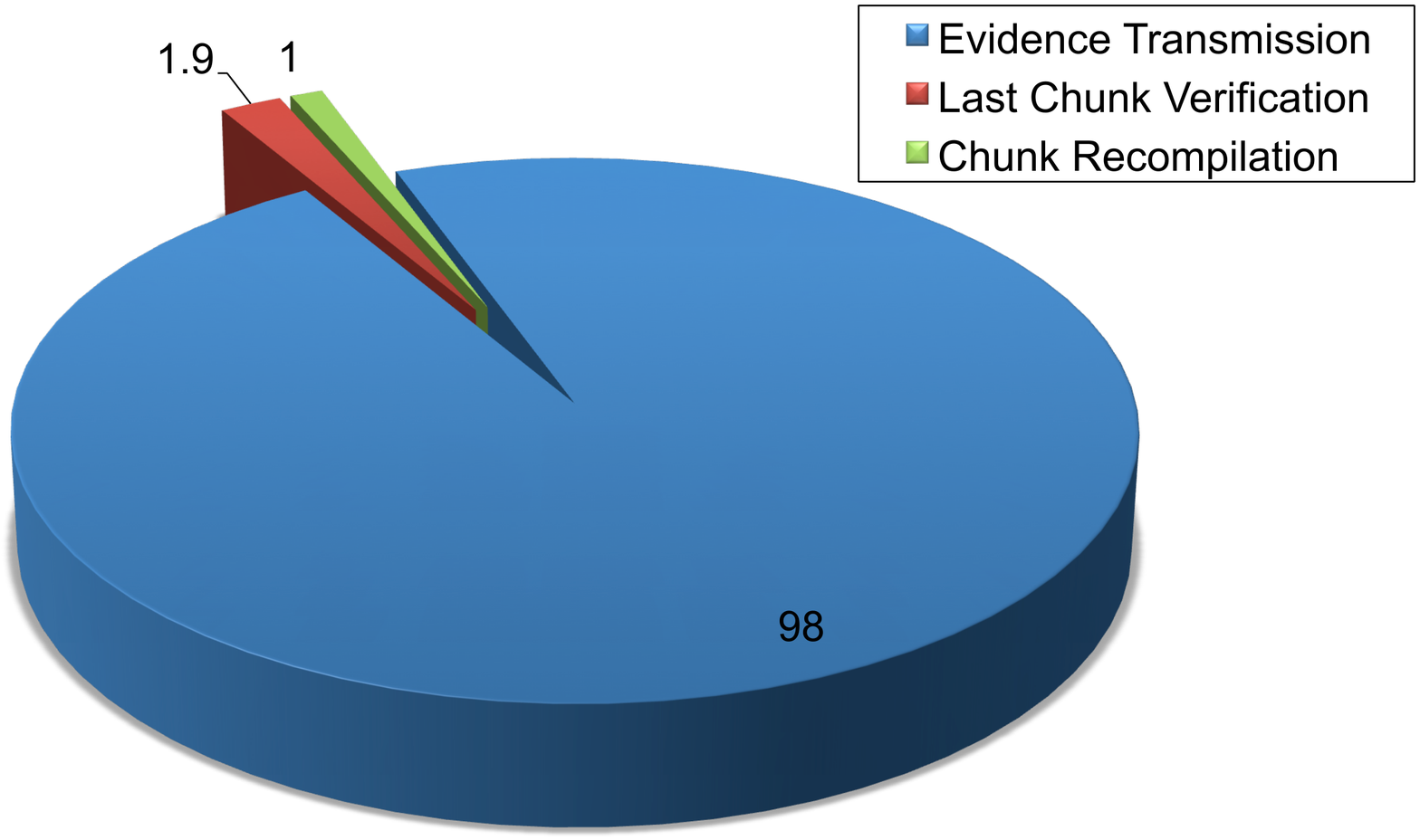}
\caption{Pie chart displaying the proportions of each of the phases in the successful acquisition of a 1000MB image (chunk size: 100MB).}
\label{fig:overalloverview}
\end{figure}

Figure \ref{fig:overalloverview} shows the result of an experiment of acquiring a 1000MB image using the RAFT system. The Internet connection used for this experiment was the same connection as used in the enterprise experiment, as outlined in section \ref{ch4:enterprise}, with a 83.26Mbps upload speed. Note that the combined overhead for the forensic verification process and the recombination process amounts to 2.9 seconds, from the overall acquisition time of 94.9 seconds (3.05\% of the total time). Due to the fact that each chunk received, once successfully verified, is then combined onto the end of the image being acquired, the time overhead for the recombination of the chunks is solely the time it takes to merge the last chunk onto the existing partially merged image.

\section{Summary}
\label{ch4:summary}
The above results from testing the viability and performance of the system described in chapter \ref{ch:arch} prove that the system is a viable tool for the remote collection of digital forensic evidence. The options available for each component were also tested in terms of speed and reliability in section \ref{ch4:viability}. This ensured that the final specification of the RAFT system has a robust architecture.

A comparison was made between the chance of a collision in commonly accepted court admissible evidence sources, i.e., a comparison was made between human DNA and fingerprinting and the evidence collected as part of the RAFT system. It was proven that due to the use of the SHA-512 hashing algorithm, that the chance of a collision being found in the evidence collected using RAFT is over $10^{113}$ times less likely than a collision being found in human DNA evidence and more than $10^{147}$ times less likely than human fingerprint evidence.

\chapter{Conclusion and Future Work}
\label{ch:conclusion}

While the time taken to image a suspect computer over the Internet is substantially longer than the time taken using traditional forensic methods (with direct physical access to the hard drive \cite{ray}) the traditional approach does not factor in the time wasted by forensic professionals in the collection of this evidence. Using RAFT could give forensic investigators the power to remotely conduct investigations in more places at once.

One significant plus of using the RAFT system is that it is extremely cost effective to distribute the client side of the system over many locations. In the law enforcement scenario, this could result in every police station having a copy of the RAFT Client. This would result in granting regular police officers the ability to quickly image a suspect computer, after receiving any necessary warrants.

\section{Usage Adaptations}
\label{ch5:adaptations}
The proposed system in this thesis was designed to serve the purpose of reducing the time wasted by forensic investigators while travelling to physically collect machines from crime scenes which results in them not being able to devote enough of their time to performing the analysis of the collected evidence. Without any modification to the system, it is also ideal for additional usage scenarios whereby a verified secure copy from a digital storage device may be required.

\subsection{Local Network Forensics}
\label{ch5:local}
In order for the RAFT system to function on a local area network as opposed to the Internet, the customisation process for the system would be exactly the same as described for the Internet based system in section \ref{ch3:adaption}. The only difference is that the hard coded hostname in the RAFT client would need to be the IP address of the server on the local network, as opposed to the Internet IP address. This would route the traffic over the local networking infrastructure and avoid the Internet access requirement. 

Using the RAFT tool over a gigabit LAN connection would significantly improve the acquisition time as the bottleneck in the system would switch to the physical storage device's read speed as opposed to the Internet connection speed. This would result in RAFT acquiring forensically verified evidence in a similar amount of time as the existing acquisition tools. One significant advantage to using the RAFT system over a LAN connection when compared to using other forensic tools, is that the RAFT system is compatible with numerous hardware configurations. The RAFT system will be able to acquire evidence from any target computer with any type of storage device, e.g., IDE hard drive, SATA hard drive, Solid State Disk (SATA or soldered onto the motherboard), memory cards etc., so long as the target computer is able to boot from a CD or USB flash drive.

\subsection{Secure Verified Backup Tool}
\label{ch5:backup}
Again, without any modification to the existing system, the RAFT system could be used as a secure, verified backup tool. As a backup tool, the user will be able to choose which partition, hard drive or other storage drive they would like backed up and the RAFT system will be able to ensure them that the image was backed up and verified to be exactly the same as the original source.

\section{Further Development}
\label{ch5:further}
While the objectives of the research outlined in this thesis were met, there are some ideas and features which could be added to or used in conjunction with the existing system to improve the overall level of functionality. Potential modifications to the current system include implementing dynamic chunk sizes, remote control/remote investigation, live system acquisition and the addition of a forensically sound compression algorithm to decrease the transfer size. There are also two additional tools that would complement the RAFT system; a hardware device design to capture live evidence before rebooting into the RAFT system and combining the RAFT system with a distributed analysis system to further improve investigation time.

These modifications and complementary tools are outlined in the following subsections.

\subsection{Dynamic Chunk Sizes}
\label{ch5:dynamic}
Implementing a feature whereby the size of each chunk is calculated on-the-fly during imaging would greatly help to reduce the time lost should a chunk be unsuccessfully transmitted. This would dynamically decrease the size of each chunk for slower, more unreliable connections, and increase the size for faster, more reliable connections.

\subsection{Remote Control and Investigation}
\label{ch5:remotecontrol}
The RAFT System could be improved upon by giving optional total remote control of the RAFT Client to the digital forensic investigator (after a suspect computer is booted by a law enforcement officer). If necessary, the investigator could remotely browse files on all read-only mounted media connected to the suspect PC without the requirement to first take an entire hard drive image. This would enable the investigator to determine if the suspect computer is relevant to the current case and could help focus the investigation quickly on the computer(s) relevant to the crime being investigated. The imaging process could also be streamlined, focusing onto the necessary hard drives or partitions on the suspect computer, i.e., targeting the image acquisition to the most relevant drives first.

\subsection{Live System Evidence Acquisition}
\label{ch5:livesystem}
A complementing tool could be created to be used in conjunction with the RAFT Client with the specific purpose of dealing with imaging a live system. This tool would have the ability to collect additional evidence from a live system before rebooting and using the RAFT client, e.g., the tool could collect evidence located in memory, running process information and other system state information. While executing any program on a live system will unavoidably alter its state, this alteration would be predictable and should not interfere with the original evidence. The tool could incorporate a version of the ``memdump" *NIX command which is capable of taking a snapshot of all the information that is currently stored in memory. Carrier \cite{Carrier2006} states that while live analysis is becoming increasingly important in digital investigations, the integrity of the evidence collected may be compromised. This is due to the design of production operating systems which do not permit applications to access the kernel memory of the machine. This results in the possibility that the evidence collected from a live system could be compromised due to countermeasures employed by the user to prevent such investigation, e.g., a rootkit could be installed at the kernel level which could hide running processes or data contained in volatile memory.

\subsection{Lossless Forensic Compression}
\label{ch5:compression}
The system could also be improved upon by eliminating unnecessary transfer size, thus improving on the time required to collect all necessary evidence, e.g., through the employment of a lossless forensically sound compression algorithm. Some of the digital evidence storage formats incorporate a level of forensically sound compression. While many of these formats are closed source, an open source alternative could be employed, such as Generic Forensic Zip \cite{gfzip}.

\subsection{Specific Hardware Device}
\label{ch5:hardware}
A specific hardware device could be created to connect to a suspect computer through an available USB connection. When powered with the USB bus, this device would execute an automated program on the device itself, which would be able to acquire evidence from the suspect computer via direct memory access (DMA) channels. This device could have an amount of flash memory incorporated into its design, such that a dump of all volatile memory could be stored on the device.

Such a device could be combined with the existing RAFT system. The procedure for using the system for law enforcement officers could be modified. For example, if the computer is turned on, first plug in the hardware and wait for it to complete a memory dump of the target computer. Then reboot the computer into the RAFT client as normal to collect all evidence from semi-permanent storage sources.

\subsection{Distributed Analysis}
\label{ch5:distributed}
The RAFT system could be combined with a distributed digital forensic analysis system to help further improve on the time taken to conduct each digital forensic investigation, such as the systems outlined in \cite{gao}, \cite{richard}, \cite{roussev} and \cite{kulesh}. An automated pass of the evidence collected could be configured whereby upon successful receipt of a new image, the first pass could be configured to categorise the data contained within the image, e.g., images, office files, Internet history etc.

\subsection{Usability Test}
\label{ch5:usability}
As outlined as part of the technical requirements of the RAFT system in section \ref{ch3:techrequirements}, the client side of the tool should be relatively easy to use for regular law enforcement officers and should require minimal training. In order to measure this requirement, a usability test should be conducted. The usability test should invite law enforcement officers and law enforcement digital forensic investigators to take part. The groups should be randomly divided into two teams, each given the same task of collecting all available digital evidence from a suspect computer:

\begin{enumerate}
\item One team would not be given any instruction on how to use the tool (besides instructions on how to boot the suspect computer using RAFT).
\item The second team would be given a short introduction to using the tool, how it operates and the best practices while using the tool.
\end{enumerate}

Should both teams achieve their task in a similar timeframe, the ease of use of the tool would be proven. This result would also demonstrate the lack of required digital forensic expertise to use the tool. Any feedback from the usability testing should be integrated into the current system.




\section{Conclusion}
\label{ch5:conclusion}

The phenomenon of the ever increasing number of crimes being aided by computers and Internet is set to continue into the future due to relative level of anonymity provided to Internet users. As a result of this inevitable increase in cybercrimes, digital forensic investigators' workload is set to increase exponentially. Any extra of the investigators' time that can be allocated to performing the analysis of cases will help to aid the turn around time for investigations. This thesis proposed and validated the viability of a forensically sound digital forensic evidence acquisition tool capable of being used by any law enforcement officer. The tool will get the evidence into an ``investigation-ready" state in the forensic laboratory as early into the investigation as possible. The traditional model for digital evidence acquisition requires the digital investigator to leave the forensic lab to visit the crime scene to collect the suspect machines. These machines may then lay untouched (and unimaged) in an evidence store for a prolonged period of time. During this time, potentially case critical information may lay undiscovered. The use of the RAFT system can significantly improve on this traditional model.
\singlespacing
\bibliographystyle{unsrt}
\bibliography{thesis}

\begin{thebibliography}{10}

\bibitem{gao}
Y.~Gao, G.G.~Richard III, and V.~Roussev.
\newblock Bluepipe: A scalable architecture for on-the-spot digital forensics.
\newblock {\em International Journal of Digital Evidence}, 3(1), Summer 2004.

\bibitem{carrier-open}
B.~Carrier.
\newblock Open source digital forensics tools: The legal argument.
\newblock {\em @stake Research Report}.

\bibitem{battistoni}
R.~Battistoni, A.~Di~Biagio, R.~Di~Pietro, M.~Formica, and L.V. Mancini.
\newblock A live digital forensic system for windows networks.
\newblock {\em Proceedings of The International Federation for Information
  Processing (IFIP) 23$^{rd}$ International Information Security Conference},
  278(2):88--97, 2008.

\bibitem{dcfldd}
DCFLDD~(Department of~Defence Computer Lab Dataset~Definition).
\newblock {\em http://dcfldd.sourceforge.net/}.
\newblock April 2009.

\bibitem{forte2006state}
D.~Forte.
\newblock The state of the art in digital forensics.
\newblock {\em Web Technology}, 67:253, 2006.

\bibitem{guidance}
EnCase~Forensic Features.
\newblock {\em http://www.guidancesoftware.com/WorkArea/
  DownloadAsset.aspx?id=671}.
\newblock Guidance Software, August 2009.

\bibitem{garber2001encase}
L.~Garber.
\newblock Encase: A case study in computer-forensic technology.
\newblock {\em IEEE Computer Magazine}, January 2001.

\bibitem{ftk}
Access Data's Forensics~Brochure Forensic~Toolkit.
\newblock {\em
  http://www.accessdata.com/downloads/media/AccessData\_Forensics\_Brochure.pdf}.
\newblock August 2009.

\bibitem{dfrws2006}
Common Digital Evidence Storage~Format (CDESF).
\newblock Survey of existing disk image storage formats.
\newblock In {\em Proc. Digital Forensic Research Workshop 2006}, September
  2006.

\bibitem{fred}
Digital~Intelligence Forensic Recovery~of Evidence Device~(FRED).
\newblock {\em http://www.digitalintelligence.com/products/fred/}.
\newblock August 2009.

\bibitem{freddie}
Digital~Intelligence Forensic Recovery~of Evidence Device~(FRED).
\newblock {\em http://www.digitalintelligence.com/products/freddie/}.
\newblock August 2009.

\bibitem{cdesf}
Digital Forensic Research Workshop (DFRWS) Common Digital Evidence Storage
  Format (CDESF)~Working Group.
\newblock {\em http://www.dfrws.org/CDESF/index.shtml}.
\newblock September 2006.

\bibitem{evidencestandards}
The Common Digital Evidence Storage Format~Working Group.
\newblock Standardizing digital evidence storage.
\newblock {\em Communications of the ACM}, 49(2):67--68, 2006.

\bibitem{aaf}
S.L. Garfinkel.
\newblock Aff: a new format for storing hard drive images.
\newblock 2006.

\bibitem{containers}
G.G. Richard, V.~Roussev, and L.~Marziale.
\newblock {Forensic discovery auditing of digital evidence containers}.
\newblock {\em Digital Investigation}, 4(2):88--97, 2007.

\bibitem{gfzip}
Generic~Forensic Zip.
\newblock {\em http://www.nongnu.org/gfzip/}.
\newblock April 2009.

\bibitem{debq}
P.~Turner.
\newblock Unification of digital evidence from disparate sources (digital
  evidence bags).
\newblock {\em Digital Investigation}, 2(3):223--228, 2005.

\bibitem{debwet}
Chet Hosmer.
\newblock Digital evidence bag.
\newblock {\em Commun. ACM}, 49(2):69--70, 2006.

\bibitem{forensicallysound}
E.~Casey.
\newblock What does forensically sound really mean.
\newblock {\em Digital Investigation}, 4(2):49--50, 2007.

\bibitem{chf}
B.~Preneel.
\newblock {\em Analysis and design of cryptographic hash functions}.
\newblock PhD thesis, 1993.

\bibitem{wangmd5}
X.~Wang and H.~Yu.
\newblock How to break md5 and other hash functions.
\newblock {\em Advances in Cryptology --- EUROCRYPT 2005, Lecture Notes in
  Computer Science}, 3494:19--35, 2005.

\bibitem{certmd5}
United States Computer Emergency Readiness~Team (US-CERT).
\newblock {\em Vulnerability Note 836068
  http://www.kb.cert.org/vuls/id/836068/}.
\newblock December 2008.

\bibitem{zhang}
X.-M. Zhang and Y.~Zheng.
\newblock Gac - the criterion for global avalanche characteristics of
  cryptographic functions.
\newblock {\em Journal of Universal Computer Science}, 1(5):320--337, 1995.

\bibitem{feistel}
H.~Feistel, W.A. Notz, and J.L. Smith.
\newblock Some cryptographic techniques for machine-to-machine data
  communications.
\newblock {\em Proceedings of the IEEE}, 63(11):1545--1554, Nov. 1975.

\bibitem{wangsha1}
X.~Wang, Y.~L. Yin, and H.~Yu.
\newblock Finding collisions in the full sha-1.
\newblock {\em Advances in Cryptology --- CRYPTO 2005, Lecture Notes in
  Computer Science}, 3621:17--36, 2005.

\bibitem{gilbert}
H.~Gilbert and H.~Handschuh.
\newblock Security analysis of sha-256 and sisters.
\newblock {\em Selected Areas in Cryptography, Lecture Notes in Computer
  Science}, 3006:175--193, 2004.

\bibitem{sha3}
National~Institute of~Standards and Technology (NIST).
\newblock {\em SHA-3 Second Round Candidates}.
\newblock August 2009.

\bibitem{commons}
Science and Technology Comittee.
\newblock Forensic science on trial.
\newblock pages 75--76, 2005.

\bibitem{daubert}
Daubert v. Merrell Dow Pharmaceutils~Syllabus Supreme Court of~the
  United~States.
\newblock {\em http://supct.law.cornell.edu/supct/html/92-102.ZS.html/}.
\newblock June 1993.

\bibitem{cftt}
United State National Institute of~Standards Computer Forensic Tool
  Testing~program and Technology.
\newblock {\em http://www.cftt.nist.gov/}.
\newblock August 2009.

\bibitem{mooreslaw}
Robert~R. Schaller.
\newblock Moore's law: past, present, and future.
\newblock {\em IEEE Spectr.}, 34(6):52--59, 1997.

\bibitem{ubuntu}
Ubuntu 8.04.
\newblock {\em http://www.ubuntu.com/}.
\newblock April 2009.

\bibitem{openssh}
OpenSSH 5.2.
\newblock {\em http://www.openssh.com/}.
\newblock April 2009.

\bibitem{sshfs}
SSH Filesystem.
\newblock {\em http://fuse.sourceforge.net/sshfs.html}.
\newblock April 2009.

\bibitem{3g}
E.~Dahlman, H.~Ekstrom, A.~Furuskar, Y.~Jading, J.~Karlsson, M.~Lundevall, and
  S.~Parkvall.
\newblock The 3g long-term evolution - radio interface concepts and performance
  evaluation.
\newblock In {\em Proc. VTC 2006-Spring Vehicular Technology Conference IEEE
  63rd}, volume~1, pages 137--141, May 7--10, 2006.

\bibitem{4g}
J.~Govil.
\newblock An empirical feasibility study of 4gs key technologies.
\newblock In {\em Proc. IEEE International Conference on Electro/Information
  Technology EIT 2008}, pages 267--270, May 18--20, 2008.

\bibitem{wang}
S-J. Wang.
\newblock Measures of retaining digital evidence to prosecute computer-based
  cyber-crimes.
\newblock {\em Computer Standards \& Interfaces}, 29(2):216--223, Febuary 2007.

\bibitem{scanlon}
M.~Scanlon and M-T. Kechadi.
\newblock Online acquisition of digital forensic evidence.
\newblock In {\em Proceedings International Conference on Digital Forensics and
  Cyber Crime (ICDF2C 2009)}, Albany, New York, USA, September 2009. Elsevier
  Limited.

\bibitem{richard}
Golden~G. Richard, III and Vassil Roussev.
\newblock Next-generation digital forensics.
\newblock {\em Communications of the ACM}, 49(2):76--80, 2006.

\bibitem{roussev}
V.~Roussev and G.G.~Richard III.
\newblock Breaking the performance wall: The case for distributed digital
  forensics.
\newblock In {\em Proceedings of the 2004 Digital Forensics Research Workshop},
  Baltimore, Maryland, USA, August 2004. DFRWS.

\bibitem{kulesh}
Kulesh Shanmugasundaram.
\newblock {\em Fornet: a distributed forensics network}.
\newblock PhD thesis, Brooklyn, NY, USA, 2006.
\newblock Adviser-Memon, Nasir.

\bibitem{dna}
D.H. Kaye.
\newblock Probability, individualization, and uniqueness in forensic science
  evidence: Listening to the academies.
\newblock {\em Brooklyn Law Review, Forthcoming}, June 2006.

\bibitem{fingerprint}
S.~Pankanti, S.~Prabhakar, and A.K. Jain.
\newblock On the individuality of fingerprints.
\newblock {\em IEEE Transactions on Pattern Analysis and Machine Intelligence},
  pages 1010--1025, 2002.

\bibitem{speedtest}
Speedtest.
\newblock {\em http://www.speedtest.net/mini.php/}.
\newblock April 2009.

\bibitem{ray}
Indrajit Ray and Sujeet Shenoi.
\newblock {\em Advances in Digital Forensics IV}.
\newblock Springer Publishing Company, Incorporated, 2008.

\bibitem{Carrier2006}
B.~D. Carrier.
\newblock Risks of live digital forensic analysis.
\newblock {\em Communications of the ACM}, 49(2):56--61, 2006.

\end{thebibliography}
\end{document}